\documentclass[twocolumn,seceq,graphicx]{jpsj2} 
\newcommand{\eps}{\varepsilon}

\newcommand{\vct}[1]{\mbox{\boldmath #1}}
\def\runauthor{Masahito {\sc Mochizuki} and Masatoshi {\sc Imada}}
\title{G-type antiferromagnetism and orbital ordering due to the crystal field from the rare-earth ions induced by the GdFeO$_3$-type distortion in $R$TiO$_3$ with $R=$La, Pr, Nd and Sm}
\author
{Masahito {\sc Mochizuki}$^1$ and Masatoshi {\sc Imada}$^{2,3}$}
\inst
{$^1$Department of Physics, University of Tokyo, Hongo, Tokyo
113-0033, Japan\\
$^2$Institute for Solid State Physics, University of Tokyo, 
Kashiwanoha, Kashiwa, Chiba 277-8581, Japan\\
$^3$PRESTO, Japan Science and Technology Agency}
\recdate{\today}

\abst{The origin of the antiferromagnetic order and puzzling properties of LaTiO$_3$ as well as the magnetic phase diagram of the perovskite titanates are studied theoretically. We show that in LaTiO$_3$, the $t_{2g}$ degeneracy is eventually lifted by the La cations in the GdFeO$_3$-type structure, which generates a crystal field with nearly trigonal symmetry. This allows the description of the low-energy structure of LaTiO$_3$ by a single-band Hubbard model as a good starting point. The lowest-orbital occupation in this crystal field stabilizes the AFM(G) state, and well explains the spin-wave spectrum of LaTiO$_3$ obtained by the neutron scattering experiment. The orbital-spin structures for $R$TiO$_3$ with $R$=Pr, Nd and Sm are also accounted for by the same mechanism. We point out that through generating the $R$ crystal field, the GdFeO$_3$-type distortion has a universal relevance in determining the orbital-spin structure of the perovskite compounds in competition with the Jahn-Teller mechanism, which has been overlooked in the literature. Since the GdFeO$_3$-type distortion is a universal phenomenon as is seen in a large number of perovskite compounds, this mechanism may also play important roles in other compounds of this type.}

\kword
{perovskite titanate, GdFeO$_3$-type distortion, G-type antiferromagnetism, orbital ordering, orbital degeneracy, rare-earth cation, crystal field splitting}
\begin{document}
\maketitle
\section{Introduction}

A series of the perovskite-type transition-metal oxides have attracted much interest since strongly correlated electrons in some of these compounds show astonishing transport phenomena upon carrier doping such as high $T_c$ superconductivity in layered cuprates and colossal magnetoresistance in manganites.~\cite{Imada98} Perovskite titanium oxide is also one of the typical examples. However, in spite of the structural similarity, the titanates show very different properties from the above two compounds.~\cite{Imada98}

The perovskite titanate $R_{1-x}A_{x}$TiO$_3$ ($R$=rare-earth ions, $A$=alkaline ions) shows a transition from insulator to metallic state upon carrier doping achieved by increasing $x$.~\cite{Tokura93a,Taguchi93,Kumagai93,Tokura93b,Fujimori92,Fujishima92} This metallic phase proved to be well described by the Fermi liquid picture, and the origin of the very different behavior of the titanates from the cuprates and manganites has been a subject of very intensive studies. More concretely, strong mass enhancement was found in the specific heat and the susceptibility near the metal-insulator transition point,~\cite{Kumagai93,Tokura93b} and discussed in connection to the nature of the Mott transition.~\cite{Furukawa92,Imada93,Rozenberg94} The nature of the transition shows a marked contrast with the cases in the cuprates and manganites, and the titanates have been recognized as a touchstone material in clarifying the strong electron correlation effects.

\begin{figure}[tdp]
\includegraphics[scale=0.4]{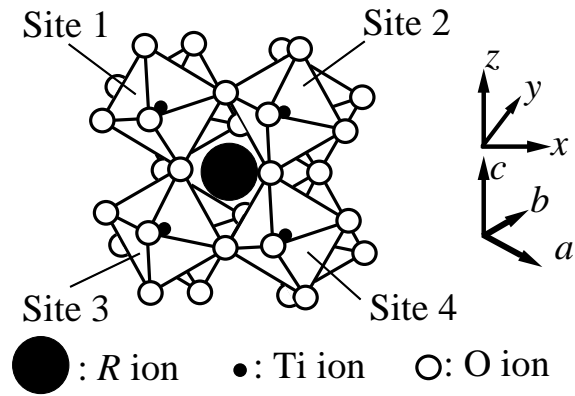}
\caption{Crystal structure of $R$TiO$_3$ with the GdFeO$_3$-type distortion.}
\label{gdfo3}
\end{figure}
In order to clarify the nature of the metallic phase and the transition, it is also essentially important to clarify the electronic structure of the end Mott-Hubbard insulator $R$TiO$_3$. In fact, $R$TiO$_3$ shows a number of puzzling and rich electronic structures arising from the orbital degrees of freedom. In these compounds, Ti$^{3+}$ has a $t_{2g}^1$ configuration in which one of the threefold $t_{2g}$ orbitals is occupied by an electron, and the 3$d$ $t_{2g}$ bands are at the Fermi level in contrast with the cuprates and manganites with $e_g$ bands at the Fermi level. One of the central puzzles in $R$TiO$_3$ is, as we will describe details below, that orbitally degenerate systems, in general, have a strong tendency towards antiferro-orbital and ferromagnetic state,~\cite{Imada98} while LaTiO$_3$ shows actually antiferromagnetic order at low temperatures. Here, ``antiferro-orbital order" is defined as an order of orbital polarization with the staggered pattern in analogy with the staggered spin polarization in the antiferromagnetic order. One of the purposes of this paper is to solve this puzzle and clarify the origin of the antiferromagnetic order. In the course of the clarification, we show that LaTiO$_3$ and related compounds are, in reality, a good prototype of the single-band Hubbard model on a cubic lattice.

The crystal structure of $R$TiO$_3$ is a pseudocubic perovskite with an orthorhombic distortion (GdFeO$_3$-type distortion) in which the TiO$_6$ octahedra forming the perovskite lattice tilt alternatingly. Note that, as a first approximation, the cubic TiO$_6$ octahedra are not atomically distorted with this distortion. In this structure, the unit cell contains four TiO$_6$ octahedra (sites 1-4) as shown in Fig.~\ref{gdfo3}. The magnitude of the distortion depends on the ionic radii of the $R$ ions. With a small ionic radius of the $R$ ion, the lattice structure is more distorted and the Ti-O-Ti bond angle is decreased more significantly from 180$^{\circ}$. For example, in LaTiO$_3$, the bond angle is 157$^{\circ}$ ($ab$-plane) and 156$^{\circ}$ ($c$-axis), but 144$^{\circ}$ ($ab$-plane) and 140$^{\circ}$ ($c$-axis) in YTiO$_3$.~\cite{MacLean79} The ionic radii of La and Y ions are 1.17 $\AA$ and 1.04 $\AA$, respectively. Here, we note that the magnitude of the distortion can also be controlled by using solid-solution systems. For example, by varying the Y concentration in La$_{1-x}$Y$_{x}$TiO$_3$, we can control the bond angle almost continuously from 157$^{\circ}$ ($x$=0) to 140$^{\circ}$ ($x$=1). In the perovskite titanates, this bond angle distortion controls the interplay of the orbital, spin and lattice degrees of freedom, and the system shows various orbital-spin phases and their phase transitions due to their interplay. It is widely recognized that the bond angle primarily controls the electron transfers between the neighboring Ti $t_{2g}$ orbitals mediated by the O $2p$ orbitals, thereby reduces the $t_{2g}$ bandwidth with decreasing bond angle. In this paper, we will show that the tilting plays another crucial role in determining the electronic structures.

\begin{figure}[tdp]
\includegraphics[scale=0.45]{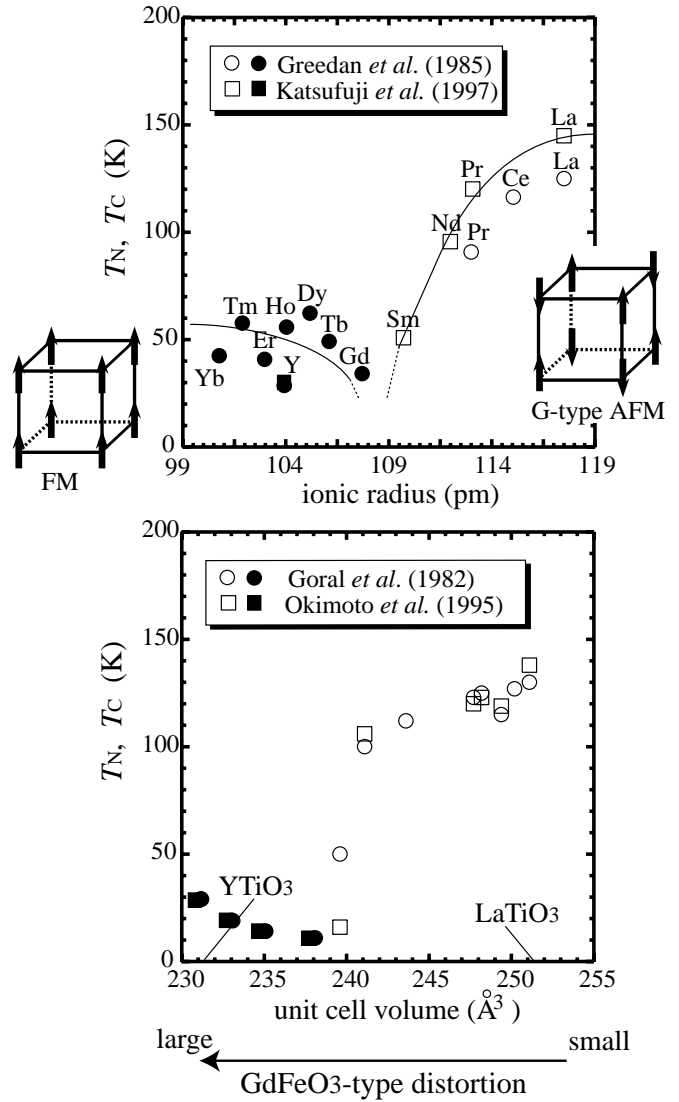}
\caption{Experimentally obtained magnetic phase diagram for $R$TiO$_3$ (upper panel, Ref.~\citen{Greedan85,Katsufuji97}), and that for La$_{1-x}$Y$_x$TiO$_3$ (lower panel, Ref.~\citen{Goral82,Okimoto95}). In these phase diagrams, $T_{\rm N}$ and $T_{\rm C}$ are plotted as functions of the ionic radius of the $R$ ion and the unit cell volume, respectively. (The lines are guides for the eyes.) Note that the unit cell volume of La$_{1-x}$Y$_x$TiO$_3$ is approximately proportional to the Y concentration ($x$), and well characterizes the magnitude of the GdFeO$_3$-type distortion.~\cite{Goral82} Here, the closed symbols and the open symbols indicate $T_{\rm N}$ and $T_{\rm C}$, respectively. The values of the unit cell volume are deduced from the data obtained by the previous x-ray diffraction measurement (Ref.~\citen{Goral82}).}
\label{mgPDG}
\end{figure}
Recently, a magnetic phase diagram in the plane of temperature and the magnitude of this distortion was obtained experimentally for $R$TiO$_3$, which exhibits an antiferromagnetic (AFM)-to-ferromagnetic (FM) phase transition (see Fig.~\ref{mgPDG} upper panel).~\cite{Greedan85,Katsufuji97} LaTiO$_3$ with the smallest distortion shows a G-type AFM (AFM(G)) ground state, in which spins are aligned antiferromagnetically in all $x$, $y$ and $z$ directions as shown in the inset of the upper panel of Fig.~\ref{mgPDG}. The magnetic moment is 0.45 $\mu_{\rm B}$,~\cite{Goral83} which is reduced from spin-1/2 moment. The N\'eel temperature ($T_{\rm N}$) is about 130 K. With increasing GdFeO$_3$-type distortion, $T_{\rm N}$ decreases and is strongly depressed at SmTiO$_3$, subsequently a FM ordering appears. In the significantly distorted compounds such as GdTiO$_3$ and YTiO$_3$, a FM ground state accompanied by a large Jahn-Teller distortion is realized. A similar phase diagram was also obtained for La$_{1-x}$Y$_{x}$TiO$_3$ (see Fig.~\ref{mgPDG} lower panel).~\cite{Goral82,Okimoto95}

A suppression of the Curie temperature ($T_{\rm C}$) around the AFM-FM phase boundary is understood from the strong two-dimensional anisotropy in the FM coupling near the transition point.~\cite{Mochizuki00,Mochizuki01a} On the other hand, in spite of a lot of attempt and effort, the origin of the AFM(G) structure in LaTiO$_3$ has been controversial. We briefly summarize the history of the study on this controversy in the following. 

Previous diffraction studies~\cite{MacLean79,Eitel86} showed that a Jahn-Teller type distortion of the TiO$_6$ octahedra in LaTiO$_3$ was undetectablly small in contrast with the cuprates and manganites. If this is true, in LaTiO$_3$, the crystal field from the O ions surrounding a Ti$^{3+}$ ion has a cubic symmetry so that the degeneracy of the $t_{2g}$ orbitals is expected to survive. In fact, a number of papers have assumed this degeneracy as we describe below, and this is a source of controversies. 

Under this circumstance, the AFM(G) ground state has been surprising and controversial since in the orbitally degenerate system, it is theoretically expected that a FM state with antiferro-orbital ordering is favored both by electron transfers and by Hund's rule coupling.~\cite{Roth66,Kugel72,Kugel73,Inagaki73,Cyrot73,Cyrot75} Indeed, a recent theoretical study based on a Kugel-Khomskii model showed that in the cubic lattice with threefold degenerate $t_{2g}$ orbitals, a FM state is almost always stable, and the AFM(G) state appears only in the unphysical regions of spin-exchange parameters.~\cite{Ishihara02}

Mizokawa and Fujimori performed a Hartree-Fock analysis of the multiband $d$-$p$ model, and proposed that the spin-orbit (LS) interaction lifts the $t_{2g}$ degeneracy, and the AFM(G) state in LaTiO$_3$ is accompanied by the LS ground state.~\cite{Mizokawa96a,Mizokawa96b} In the LS ground state, unquenched orbital moment antiparallel to the spin moment is necessarily induced. At first sight, this seems to be consistent with the experimentally observed reduced magnetic moment in this compound.

However, in contrast with this naive expectation, a recent neutron scattering study has shown a spin-wave spectrum well described by an isotropic spin-1/2 Heisenberg model ($J\sim$15.5 meV) with a considerably small spin gap.~\cite{Keimer00} This indicates that the orbital moment is almost quenched, and the LS interaction is irrelevant since if the orbital moment exists, an induced Ising-type anisotropy in the spin sector has to generate a large spin gap. Actually, an exact diagonalization study has recently shown that the Kugel-Khomskii model with LS interaction does not describe the observed spin-wave spectrum.~\cite{Miyahara02} In addition, a model Hartree-Fock study done by one of the authors~\cite{Mochizuki02} also showed that the LS ground state cannot reproduce the AFM(G) ground state by examination of a solution which was overlooked in the previous Mizokawa and Fujimori's study. In Ref.~\citen{Mochizuki02}, it was shown that even without the static Jahn-Teller distortion, a FM state out of which two states $\frac{1}{\sqrt{2}}(yz+{\rm i}zx)\uparrow$ and $xy\uparrow$ are alternating is stabilized both by the LS interaction and by the spin-orbital superexchange interaction.

To explain the neutron scattering result, a possible orbital liquid state was proposed on the basis of small orbital-exchange interaction in the AFM(G) spin structure.~\cite{Khaliullin00,Khaliullin01,Kikoin03} In these theories, the AFM(G) ordering is assumed a priori. However, in the titanates, the spins and orbitals strongly couple to each other, and both degrees of freedom cannot be determined independently. Therefore, the origin of the AFM(G) state in LaTiO$_3$ is to be clarified in a self-consistent manner. More importantly, a FM state with antiferro-orbital ordering was theoretically expected to be more stable in this system. Thus, the assumption of the AFM(G) order on this basis is hard to be justified since the relative stability of the AFM(G) to FM state has never been examined. In addition, a recent heat capacity measurement showed that the most of the low-temperature heat capacity arises from magnon contributions, which contradicts the prediction from the orbital liquid theory about the orbital contributions.~\cite{Fritsch02}

\begin{figure}[tdp]
\includegraphics[scale=0.5]{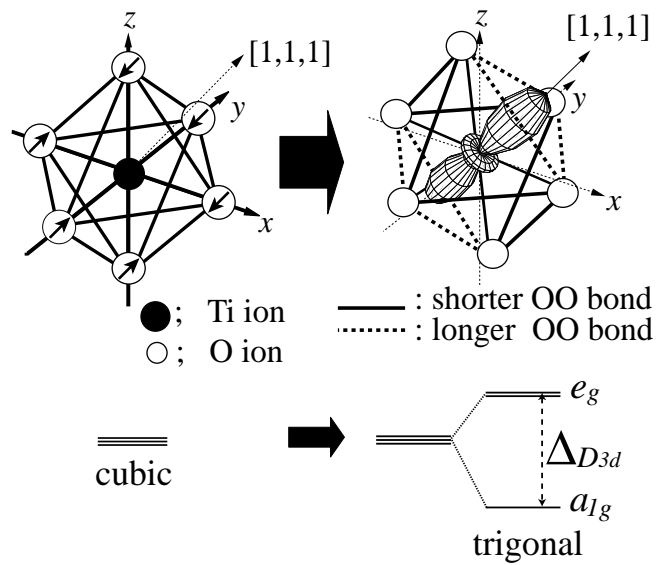}
\caption{With the $D_{3d}$ distortion, the TiO$_6$ octahedron is contracted along a trigonal direction. There exist two kinds of O-O bonds, shorter O-O bond and longer O-O bond as presented by solid lines and dashed lines, respectively. As a result of this distortion, the threefold degenerate $t_{2g}$ levels split into a nondegenerate lower $a_{1g}$ level and twofold degenerate $e_g$ levels. With the $D_{3d}$ crystal field with [1,1,1] trigonal axis, the representation of the $a_{1g}$ orbital is $\frac{1}{\sqrt{3}}(xy+yz+zx)$.}
\label{D3ddist}
\end{figure}
A possible trigonal ($D_{3d}$) distortion of the TiO$_6$ octahedra was proposed to explain the emergence of AFM(G) structure.~\cite{Mochizuki01b} With this distortion, the TiO$_6$ octahedron is contracted along the threefold direction, and the threefold degenerate $t_{2g}$ levels split into a nondegenerate lower $a_{1g}$ level and twofold-degenerate higher $e_g$ levels (see Fig.~\ref{D3ddist}). Occupations of the $a_{1g}$ orbitals well explain the emergence of the AFM(G) ordering and the isotropic spin-wave spectrum. However, this distortion had not been observed clearly upon theoretical proposal. 

NMR studies have also applied to this compound. Previously, the NMR spectrum was analyzed based on the orbital liquid picture.~\cite{Itoh97,Itoh99} However, it has recently been claimed that there exists a discrepancy. Indeed, it was recently clarified that the spectrum is well described by an orbital ordering model proposed in the above $D_{3d}$-distortion scenario instead of the orbital liquid model.~\cite{Kiyama03} Moreover, a recent resonant x-ray scattering result also indicates the orbital ordering in the series of the AFM(G) compounds of $R$TiO$_3$ ($R=$La, Pr, Nd and Sm), which has the same symmetry with that observed in the ferromagnetic YTiO$_3$ and GdTiO$_3$,~\cite{Kubota00} which also contradicts the orbital liquid picture. 

Several first-principles methods were also applied to LaTiO$_3$. However, LDA and GGA failed to reproduce not only AFM(G) structure but also insulating behavior since the strong electron correlation is insufficiently treated.~\cite{Pari95,Mahadevan96,Sawada97,Sawada98} In addition, although the insulating gap was reproduced, the AFM(G) structure was not obtained in LDA+$U$ calculation.~\cite{Solovyev96} These indicate that in this compound, the effect of the strong electron correlations is crucially important, which is of general interest of the condensed matter physics. 

In such ways as discussed above, the problem of the orbital-spin structure in LaTiO$_3$ has been studied intensively, but a consistent theory has not emerged, and is still under hot debates. In fact, when we were performing the calculations and were preparing this manuscript, several reports of new experimental findings and theoretical results appeared in the preprint server, which we will introduce and discuss later in the context.

To fully understand electron correlation effects in the perovskite compounds and in more general, in $d$-electron systems, it is recognized that puzzling properties of the Ti perovskites have to be explained by a convincing theory. The Ti perovskite compound also offers touchstone materials for the understanding of orbital physics, which is a hot topic of the physics of electron correlation.~\cite{Tokura00} Because of the threefold degeneracy of $d^1$ state in the titanates, it is expected that the complete understanding of the Ti perovskites contributes in clarifying roles of orbital degrees of freedom as well as coupling and interplay between magnetism and orbitals in the strongly correlated electron systems. The purpose of the present theoretical study is to solve the long-standing puzzles in the perovskite titanates and to offer an overall understanding of the interplay in these compounds.\\

In the perovskite compounds, the Jahn-Teller distortion often plays crucial roles in determining low-energy electronic state by lifting the orbital degeneracy. However, with early transition-metal ions, the Jahn-Teller coupling is considerably weaker than that in the late-$3d$ compounds. Indeed, in contrast with the cuprates and manganites, a sizable Jahn-Teller distortion had not been observed in LaTiO$_3$ as already mentioned above. Without the Jahn-Teller distortion, the crystal field from the ligand O ions acting on the Ti$^{3+}$ ion has a cubic symmetry.

In addition to the Jahn-Teller distortion, the GdFeO$_3$-type lattice distortion is another generic phenomenon in perovskites. However, it has been assumed mainly to control the bandwidth through the $M$-O-$M$ angle variation with $M$ being a transition-metal ion,~\cite{Okimoto95,Katsufuji95} while its direct effects on interplay of spins and orbitals have not been considered seriously. In fact, the knowledge about the symmetry lowering caused by the GdFeO$_3$-type distortion in principle could lead to the idea for the lift of the $t_{2g}$-orbital degeneracy. However, in contrast with the Jahn-Teller distortion, a unit TiO$_6$ octahedron as well as its cubic symmetry are not distorted in the generic GdFeO$_3$-type distortion so that it has been assumed in most of the literature that the threefold degeneracy of the $t_{2g}$ orbitals is retained even with this distortion. Actually, the recent orbital physics is usually based on the degenerate $t_{2g}$ and $e_g$ orbitals unless the Jahn-Teller type octahedral distortion exists. However, this assumption is in fact a source of the puzzles recognized in LaTiO$_3$.

To solve the puzzles and controversies in the titanates, we consider that it is important to reveal how the basic degeneracy of the orbitals is retained or split in the realistic experimental situation since this is a starting point of studies on orbital physics for perovskite compounds. We believe that it is crucial to establish the precise estimate about the degree of orbital degeneracy under the existence of the GdFeO$_3$-type distortion while its effects have never been considered seriously. 

In this paper, we clarify that the generic GdFeO$_3$-type distortion actually generates a new mechanism for control of low-energy orbital-spin structure through lifting of the orbital degeneracy by generating crystal fields of the $R$ ions. This mechanism competes with that of the Jahn-Teller distortion and the LS interaction.

When the Jahn-Teller distortion of the TiO$_6$ octahedra is small, we expect that the $R$ crystal field plays a dominant role on the electronic structure in the perovskites by lifting the $t_{2g}$ degeneracy. By utilizing the positional parameters of the La ions measured by the x-ray diffraction study,~\cite{MacLean79} we construct the Hamiltonian for the crystal field from the La ions in LaTiO$_3$. Analyses of the obtained Hamiltonian show that the GdFeO$_3$-type distortion, and resulting displacements of the La ions turn out to generate a crystal field with nearly trigonal symmetry. As a result, the threefold degenerate cubic-$t_{2g}$ levels split into three nondegenerate levels. Further, we study the stability and properties of the AFM(G) state in this crystal field by using an effective spin-pseudospin Hamiltonian constructed through the second-order perturbational expansion in terms of the electron transfers in the limit of the strong Coulomb repulsion. The calculations of energies and spin-exchange constant based on this effective Hamiltonian show that lifting of the $t_{2g}$ degeneracy and the lowest-orbital occupation well explain the emergence and properties of the AFM(G) state in LaTiO$_3$.

We also discuss a recent experimental finding of the octahedral distortion in LaTiO$_3$. As already mentioned above, we previously predicted the $D_{3d}$ distortion of the TiO$_6$ octahedra in LaTiO$_3$ as an origin of the AFM(G) ordering.~\cite{Mochizuki01b} After this theoretical proposal, several diffraction studies have been done,~\cite{Cwik03,Hemberger03,Arao02} and recently, the distortion was actually detected by Cwik. $et$ $al.$ (Ref.~\citen{Cwik03}). We discuss that the emergence of this distortion is naturally explained by our theory, where the crystal field from the displaced $R$ ions drives the distortion of the TiO$_6$ octahedron. We also show that in spite of this distortion, our approach in which the crystal field from the $R$ ions is treated as a perturbation to the $t_{2g}$ degeneracy is justified.

Further, in order to clarify the relation between the GdFeO$_3$-type distortion and the crystal field of the $R$ ions, we also study the crystal field Hamiltonians for $R$TiO$_3$ with $R$ being Pr, Nd and Sm. Analyses of the crystal field Hamiltonians again reproduce the AFM(G) state in each compound. The calculated spin-exchange constants well explain the observed reduction of $T_{\rm N}$ as the GdFeO$_3$-type distortion increases. Results of a recent resonant x-ray scattering experiment are also discussed by analyzing the orbital structures in detail. In addition, we also show that in SmTiO$_3$, the $R$ crystal field caused by the GdFeO$_3$-type distortion strongly competes with the O crystal field due to the Jahn-Teller type TiO$_6$ distortion. The orbital-spin structure and the depression of $T_{\rm N}$ in SmTiO$_3$ are discussed by considering this competition.

We point out the importance of the GdFeO$_3$-type distortion as a universal control mechanism of orbital-spin structure through generating the $R$ crystal field in the perovskite compounds, which has been so far overlooked. This mechanism is substantially different from a usual Jahn-Teller mechanism.

The organization of this paper is as follows. In Sec. 2, we construct the effective spin-pseudospin Hamiltonian of the perovskite Ti oxides by following an approach similar to the Kugel-Khomskii formulation.~\cite{Kugel72,Kugel73} In Sec. 3, we examine the crystal field Hamiltonian of the La ions in LaTiO$_3$. We discuss the origin and nature of the AFM(G) state as well as several experimental results for LaTiO$_3$. In particular, we focus on the spin-wave spectrum obtained by a neutron scattering study in detail. In Sec. 4, we examine the crystal field of the $R$ cations in $R$TiO$_3$ with $R$ being Pr, Nd, and Sm. On the basis of this analysis, we discuss the behavior of $T_{\rm N}$ in the magnetic phase diagram and recent results of a resonant x-ray scattering. In Sec. 5, we discuss the following five points upon our proposal: (1) Available experimental results for the titanates are well explained within our theory. (2) The mechanism proposed in this paper is novel and universal, which controls the orbital-spin structure in the perovskites, and is completely different from the usual Jahn-Teller one. (3) Numerical results here obtained are sufficiently accurate even though they are calculated based on a point charge model with the nearest neighbor ions. Further, the predicted isotropy of the AFM(G) coupling as well as the properties of the orbital-spin states in $R$TiO$_3$ are robust. (4) The experimentally obtained magnetic phase diagram is well explained in the light of the present results together with the previous ones. (5) The experimentally observed reduction of the magnetic moment in LaTiO$_3$ is quantitatively understood when we consider a large amount of itinerant fluctuations in this compound. Section 6 is devoted to the summary. A short report on a part of the content of this paper has already been published.~\cite{Mochizuki03} In this paper, we describe a comprehensive framework and results on this issue together with additional new results and more detailed discussions.

\section{Microscopic Derivation of the Effective Spin-Pseudospin Hamiltonian}

In this section, we derive the effective spin-pseudospin Hamiltonian to describe the realistic system of the perovskite titanates in the limit of the strong Coulomb repulsion. We start with the multiband $d$-$p$ model in which the full degeneracies of the Ti $3d$ and O $2p$ orbitals as well as the on-site Coulomb and exchange interactions are taken into account. The Hamiltonian is given by
\begin{equation}
H^{dp}=H_{d0}+H_{p}+H_{tdp}+H_{tpp}+H_{\rm on-site},\\
\label{dph}
\end{equation}
with
\begin{eqnarray}
  & &H_{d0}=\sum_{i,\gamma,\sigma} \eps_{d}^0
        d_{i\gamma\sigma}^{\dagger} d_{i\gamma\sigma}, \\
  & &H_{p}=\sum_{j,l,\sigma} \eps_{p}
        p_{jl\sigma}^{\dagger} p_{jl\sigma}, \\
  & &H_{tdp}=\sum_{i,\gamma,j,l,\sigma} 
        t_{i\gamma,jl}^{dp}d_{i\gamma\sigma}^{\dagger} 
                           p_{jl\sigma} + \vct{h.c.}, \\
  & &H_{tpp}=\sum_{j,l,j',l',\sigma} 
        t_{jl,j'l'}^{pp}p_{jl\sigma}^{\dagger} 
                        p_{j'l'\sigma} + \vct{h.c.}, \\
  & &H_{\rm on-site}=H_{u}+H_{u'}+H_j+H_{j'},
\label{dph2}
\end{eqnarray} 
where $d_{i\gamma\sigma}^{\dagger}$ is a creation operator of an electron with spin $\sigma$(=$\uparrow,\downarrow$) in the $3d$ orbital $\gamma$ at Ti site $i$, and $p_{jl\sigma}^{\dagger}$ is a creation operator of an electron with spin $\sigma$(=$\uparrow,\downarrow$) in the $2p$ orbital $l$ at oxygen site $j$. Here, we choose the $xy$, $yz$, $zx$, $3z^2-r^2$ and $x^2-y^2$ orbitals as the basis of the $3d$ orbitals. $H_{d0}$ and $H_p$ express the bare level energies of the Ti 3$d$ and O 2$p$ orbitals, respectively. $H_{tdp}$ and $H_{tpp}$ express the $d$-$p$ and $p$-$p$ transfers, respectively. $t_{i\gamma,jl}^{dp}$ and $t_{jl,j'l'}^{pp}$ are the nearest-neighbor $d$-$p$ and $p$-$p$ transfers given in terms of Slater-Koster parameters $V_{pd\pi}$, $V_{pd\sigma}$, $V_{pp\pi}$ and $V_{pp\sigma}$.~\cite{Slater54,Harrison89} The term $H_{\rm on-site}$ represents the on-site $d$-$d$ Coulomb interactions, which consists of the following four contributions:
\begin{eqnarray}
 & &H_{u} = \sum_{i,\gamma} u
        d_{i\gamma\uparrow}^{\dagger} d_{i\gamma\uparrow}
        d_{i\gamma\downarrow}^{\dagger} d_{i\gamma\downarrow},\\
 & &H_{u'} = \sum_{i,\gamma>\gamma',{\sigma},{\sigma}'} u'
        d_{i\gamma\sigma}^{\dagger} d_{i\gamma\sigma}
        d_{i\gamma'\sigma'}^{\dagger} d_{i\gamma'\sigma'},\\
 & &H_{j} = \sum_{i,\gamma>\gamma'\sigma,\sigma'} j
        d_{i\gamma\sigma}^{\dagger} d_{i\gamma'\sigma}
        d_{i\gamma'\sigma'}^{\dagger} d_{i\gamma\sigma'},\\
 & &H_{j'} = \sum_{i,\gamma \ne \gamma'} j'
        d_{i\gamma\uparrow}^{\dagger} d_{i\gamma'\uparrow}
        d_{i\gamma\downarrow}^{\dagger} d_{i\gamma'\downarrow},
\end{eqnarray}
where $H_{u}$ and $H_{u'}$ are the intra- and inter-orbital Coulomb interactions, respectively, and $H_{j}$ and $H_{j'}$ denote the exchange interactions. The term $H_{j}$ is the origin of the Hund's rule coupling. The term $H_{j'}$ gives the $\uparrow\downarrow$-pair hopping between the $3d$ orbitals on the same Ti atom. These interactions are expressed using Kanamori parameters, $u$, $u'$, $j$ and $j'$ which satisfy the following relations:~\cite{Brandow77,Kanamori63} $u=U+\frac{20}{9}j$, $u'=u-2j$ and $j=j'$. Here, $U$ gives the magnitude of the multiplet-averaged $d$-$d$ Coulomb interaction. The charge-transfer energy $\Delta$, which describes the energy difference between the occupied O 2$p$ level and the unoccupied Ti $3d$ level, is defined by using $U$ and the energies of the bare Ti $3d$ and O $2p$ orbitals $\eps_d^0$ and $\eps_p$ as $\Delta=\eps_{d}^0+U-\eps_p$ since the characteristic unoccupied $3d$ level energy at the singly-occupied Ti site is $\eps_{d}^0$+$U$. The values of $\Delta$, $U$ and $V_{pd\sigma}$ are estimated by the analyses of the LDA band calculation~\cite{Mahadevan96} and the cluster-model analyses of the valence-band and transition-metal $2p$ core-level photoemission spectra.~\cite{Saitoh95,Bocquet96} In the present $d$-$p$ Hamiltonian, we have ignored the Coulomb interaction between two electrons on the O $2p$ orbitals as well as those on the O $2p$ and the Ti $3d$ orbitals. We take the values of these parameters as $\Delta=5.5$ eV, $U=4.0$ eV, $V_{pd\sigma}=-2.4$ eV, $V_{pd\pi}=1.33$ eV, $V_{pp\sigma}=0.52$ eV, $V_{pp\pi}=-0.15$ eV and $j=0.46$ eV. 

In the path-integral formalism, the expression of the partition function is given by
\begin{equation}
Z = \int {\cal D} {\bar{d}}_{i\gamma\sigma}
         {\cal D} d_{i\gamma\sigma} 
         {\cal D} {\bar{p}}_{jl\sigma}
         {\cal D} p_{jl\sigma} 
\exp \left[- \int_0^{\beta} {\rm d}\tau L(\tau) \right],\\
\end{equation}
with
\begin{eqnarray}
  L(\tau ) = H^{dp}(\tau)
&+& \sum_{i,\gamma,\sigma} {\bar{d}}_{i\gamma\sigma}(\partial_\tau-\mu)
d_{i\gamma\sigma} \nonumber \\       
&+& \sum_{j,l,\sigma} {\bar{p}}_{jl\sigma}
(\partial_\tau-\mu)p_{jl\sigma},
\end{eqnarray}
where $\tau$ denotes the imaginary time introduced in the path-integral formalism, and ${\bar{d}}_{i\gamma\sigma}$, $d_{i\gamma\sigma}$, ${\bar{p}}_{jl\sigma}$ and $p_{jl\sigma}$ are the Grassman variables corresponding to the operators $d_{i\gamma\sigma}^{\dagger}$, $d_{i\gamma\sigma}$, $p_{jl\sigma}^{\dagger}$ and $p_{jl\sigma}$, respectively. By using the Matsubara-frequency representation;
\begin{eqnarray}               
     d_{i\gamma\sigma}(\tau) = \frac{1}{\sqrt{\beta}}
\sum_{\omega_n} d_{i\gamma\sigma}(\omega_n) 
{\rm e}^{-{\rm i}\omega_{n}\tau}, \\
     p_{jl\sigma}(\tau) = \frac{1}{\sqrt{\beta}}
\sum_{\omega_n} p_{jl\sigma}(\omega_n)
{\rm e}^{-{\rm i}\omega_{n}\tau},
\end{eqnarray}
we have
\begin{eqnarray}
Z = \int{\cal D}& &{\bar{d}}_{i\gamma\sigma}(\omega_n)
        {\cal D} d_{i\gamma\sigma}(\omega_n) 
        {\cal D} {\bar{p}}_{jl\sigma(\omega_n)}
        {\cal D} p_{jl\sigma}(\omega_n) \nonumber \\
  & &\times\exp \left[- \sum_{\omega_n}  L(\omega_n) \right]
\end{eqnarray}
with
\begin{eqnarray}
         L(\omega_n) &=& \sum_{i,\gamma,\sigma} 
{\bar{d}}_{i\gamma\sigma}(-{\rm i}\omega_n+\eps_d^0-\mu)
d_{i\gamma\sigma} \nonumber \\      
& &  + \sum_{j,l,\sigma} 
{\bar{p}}_{jl\sigma}(-{\rm i}\omega_n+\eps_p-\mu)
p_{jl\sigma} \nonumber \\
& &  + \sum_{i,\gamma,j,l,\sigma}t_{i\gamma,jl}^{dp} 
{\bar{d}}_{i\gamma\sigma}p_{jl\sigma}  + \vct{c.c.} \nonumber \\
& &  +  \sum_{j,l,j',l'\sigma}t_{jl,j'l'}^{pp}
{\bar{p}}_{jl\sigma}p_{j'l'\sigma}  + \vct{c.c.} \nonumber \\
& &  + H_{\rm on-site}. 
\end{eqnarray}
After integrating over the ${\bar{p}}$ and $p$, the partition function is rewritten as
\begin{equation}
Z = \int {\cal D} {\bar{d}}_{i\gamma\sigma}(\omega_n)
         {\cal D} d_{i\gamma\sigma}(\omega_n)
\exp \left[- \sum_{\omega_n} L_d(\omega_n) \right],
\end{equation}
where
\begin{equation}
L_d(\omega_n) =L_{d1}(\omega_n)+L_{d2}(\omega_n)
\end{equation}
with
\begin{eqnarray}
L_{d1}(\omega_n)&=&\sum_{i,\gamma,\sigma}
{\bar{d}}_{i\gamma\sigma}(-{\rm i}\omega_n+\eps_d^0-\mu)d_{i\gamma\sigma}
\nonumber\\
& &+H_{\rm on-site},\\
L_{d2}(\omega_n)&=&
\sum_{i,\gamma,i',\gamma',\sigma}\sum_{j,l,j',l'}
{\bar{d}}_{i\gamma\sigma}\noindent\\
\times& &\left[ t_{i\gamma,jl}^{dp} (H^{-1}_{jl,j'l'}({\rm i}\omega_n))
t_{i'\gamma',j'l'}^{dp} \right]  
d_{i'\gamma'\sigma}.
\end{eqnarray}
Here, the matrix $H_{jl,j'l'}({\rm i}\omega_n)$ takes the form
\begin{equation}
H_{jl,j'l'}(i\omega_n) 
= -(-{\rm i}\omega_n+(\eps_p-\mu)) \delta_{jl;j'l'} - t_{jl,j'l'}^{pp}.\\
\end{equation}

Since the energy scale of electron transfers $t_{pd}$ ($\sim$1.3 eV) and $t_{pp}$ ($\sim$0.3 eV) are relatively smaller than the characteristic energy of the on-site interaction $U$ (=4.0 eV), it may be justified to treat the term $L_{d2}(\omega_n)$ as a perturbation to the $H_{\rm on-site}$ term. Therefore, we approximate ${\rm i}\omega_n$ in $L_{d2}(\omega_n)$ with poles of $L_{d1}(\omega_n)$, in which dynamics of the O $2p$ orbitals are integrated out. As a result, we obtain the expression of the effective supertransfer matrix element between the Ti $3d$ orbitals, and the expression of the crystal field from the ligand oxygens acting on the Ti $3d$ orbitals as follows;
\begin{equation}
t_{i\gamma,i'\gamma'}^{dd} = 
\sum_{j,l,j',l'}H^{-1}_{jl,j'l'}(\eps_{d}^0+U-\mu)
t_{i\gamma,jl}^{dp}t_{i'\gamma',j'l'}^{dp},
\end{equation}
\begin{equation}
\eps_{d\,i,\gamma\gamma'} = \eps_{d}^0 \delta_{\gamma\gamma'}
+ \sum_{j,l,j',l'}H^{-1}_{jl,j'l'}
(\eps_{d}^0+U-\mu)
t_{i\gamma,jl}^{dp}t_{i\gamma',j'l'}^{dp}.
\end{equation}
The expressions of the indirect $d$-$d$ transfer integrals and the $3d$-level energies thus obtained are equivalent to those obtained through the higher-order perturbational expansion with respect to the $d$-$p$ and $p$-$p$ transfers whose perturbational processes contain not only one O $2p$ state but also more than one O $2p$ states. 

As a result, the "effective" multiband Hubbard Hamiltonian derived from the multiband $d$-$p$ model has the form;
\begin{equation}
  H^{\rm mH}=H_{d}^{\rm mH}+H_{tdd}^{\rm mH}+H_{\rm on-site},
\end{equation}
with
\begin{eqnarray}
     & &H_{d}^{\rm mH} = \sum_{i,\gamma,\gamma'\sigma} 
                         \eps_{d\,i,\gamma\gamma'}
        d_{i\gamma\sigma}^{\dagger} d_{i\gamma'\sigma}, \\
     & &H_{tdd}^{\rm mH} = \sum_{i,\gamma,i',\gamma',\sigma} 
        t_{i\gamma,i'\gamma'}^{dd}
                      d_{i\gamma\sigma}^{\dagger} 
                      d_{i'\gamma'\sigma}  + \vct{h.c.}, \\
     & &H_{\rm on-site} = H_{u} + H_{u'} + 
                          H_j + H_{j'}, 
\label{mhh}
\end{eqnarray} 
Here, $H_{d}^{\rm mH}$ expresses the energy levels of the $3d$ orbitals in the crystal field from the ligand oxygens. $H_{tdd}^{\rm mH}$ is the $d$-$d$ supertransfer term. 

The advantages of the above formalism are as follows:
\begin{itemize}
\item In the path-integral formalism, we can evaluate the amplitudes of the $d$-$d$ transfers by treating the perturbational processes which contain more than one O $2p$ states also as the intermediate states systematically.
\item The obtained formulas of the $d$-$d$ transfer and the crystal field terms are expressed by using parameters of the multiband $d$-$p$ model. Since the values of these parameters have been well evaluated both theoretically and experimentally, we can estimate the amplitudes of the $d$-$d$ transfer integrals and the magnitude of the crystal field quantitatively.
\end{itemize}

By applying the second-order perturbational expansion to thus obtained multiband Hubbard Hamiltonian, we derive an effective Hamiltonian on the subspace of states only with singly-occupied $t_{2g}$ orbitals at each Ti site in the insulating limit. We carry out the following expansion in terms of $d$-$d$ transfers and $j/U$:
\begin{eqnarray}
 H_{\rm eff}^{(2)} &=& H_{tdd}^{\rm mH}
 \frac{1}{E_0-(H_{u+u'}+H_{j+j'})} H_{tdd}^{\rm mH}    \nonumber\\
 &=& H_{tdd}^{\rm mH}\frac{1}{E_0-H_{u+u'}} \Bigg(1    \nonumber\\
 & &+\frac{1}{E_0-H_{u+u'}}H_{j+j'}                    \nonumber\\
 & &+\frac{1}{E_0-H_{u+u'}}H_{j+j'}\frac{1}{E_0-H_{u+u'}}
 H_{j+j'}                                              \nonumber\\
 & &+ ..... \Bigg)  H_{tdd}^{\rm mH},
\end{eqnarray}
where
\begin{eqnarray}
      & &H_{u+u'} = H_{u} + H_{u'},\\
      & &H_{j+j'} = H_{j} + H_{j'}.
\end{eqnarray}

When one of the threefold $t_{2g}$ orbitals is occupied at each site, we can describe the electronic state using a spin-1 operator, which we call the pseudospin $\vct{$\tau$}$. We express the occupied $xy$ orbital by a quantum number ${\tau}_z = -1$, the $yz$ orbital by ${\tau}_z = 0$ and the $zx$ orbital by ${\tau}_z = +1$. In this subspace, we rewrite the $3d$ electron operator ${d^{\dagger}}_{i m' \sigma'} d_{i m \sigma}$ as $A_{m,m'}B_{\sigma,\sigma'}$ in terms of ${\vct{$S$}}$ and ${\vct{$\tau$}}$ depending on the indices $m$, $m'$, $\sigma$ and $\sigma'$. Here, the indices $m$ and $m'$ run over the $xy$ $yz$ and $zx$ orbitals. The symbol $\sigma$ and $\sigma'$ denote the spin of the $3d$ electron. The definitions of $A_{m,m^{\prime}}$ and $B_{\sigma,\sigma'}$ are as follows,
\begin{eqnarray}
& &A_{m,m'}=  \nonumber \\
& &\left\{
\begin{array}{ll}
 1/2\quad {\tau_z}({\tau_z} - 1)
  \equiv\beta^{-1}  &\mbox{for}\quad (m,m')=(1,1) \\
 1 - {\tau_z^2}
  \equiv\beta^{0}   &\mbox{for}\quad (m,m')=(2,2) \\ 
 1/2\quad {\tau_z}({\tau_z} + 1)
  \equiv\beta^{+1}  &\mbox{for}\quad (m,m')=(3,3) \\  
 1/\sqrt2\quad {\tau^+}\beta^{-1}  &\mbox{for}\quad (m,m')=(1,2) \\ 
 1/2\quad {\tau^+}{\tau^+}         &\mbox{for}\quad (m,m')=(1,3) \\ 
 1/\sqrt2\quad {\tau^-}\beta^{0}   &\mbox{for}\quad (m,m')=(2,1) \\ 
 1/\sqrt2\quad {\tau^+}\beta^{0}   &\mbox{for}\quad (m,m')=(2,3) \\ 
 1/2\quad {\tau^-}{\tau^-}         &\mbox{for}\quad (m,m')=(3,1) \\ 
 1/\sqrt2\quad {\tau^-}\beta^{+1}  &\mbox{for}\quad (m,m')=(3,2) \\ 
\end{array} \right. \nonumber \\
\\
\nonumber \\
& &B_{\sigma,\sigma'}=  \nonumber \\
& &\left\{
\begin{array}{ll}
 1/2\; + S_z  &\mbox{for}\quad (\sigma,\sigma')=(\uparrow,\uparrow)     \\
 1/2\; - S_z  &\mbox{for}\quad (\sigma,\sigma')=(\downarrow,\downarrow) \\
 S^+          &\mbox{for}\quad (\sigma,\sigma')=(\uparrow,\downarrow)   \\
 S^-          &\mbox{for}\quad (\sigma,\sigma')=(\downarrow,\uparrow),
\end{array} \right. 
\end{eqnarray}
where
\begin{eqnarray}
{\tau^+}={\tau_x}+{\rm i}{\tau_y},\quad
{\tau^-}={\tau_x}-{\rm i}{\tau_y}, 
\end{eqnarray}
and
\begin{eqnarray}
S^+ = S_x + {\rm i}S_y,\quad
S^- = S_x - {\rm i}S_y.
\end{eqnarray}
Here, the numbers $m$ ($m'$)=1, 2, and 3 correspond to the $xy$, $yz$ and $zx$ orbitals, respectively. 

We express the $3d$ electron operators in terms of $\vct{$S$}$ and $\vct{$\tau$}$ in accordance with the above rules to arrive at the effective spin-pseudospin Hamiltonian;
\begin{equation}
  H_{\rm eff} = \tilde {H}_{d}^{\rm mH} + H_{s\tau}.
\end{equation}
The first term $\tilde {H}_{d}^{\rm mH}$ denotes the crystal field from the O ions acting on the Ti $t_{2g}$ orbitals. Without any distortion of the TiO$_6$ octahedron, the crystal field ($\tilde {H}_{d}^{\rm mH}$) has a cubic symmetry so that it does not lift the $t_{2g}$ degeneracy. In the following calculations, in order to study the effects of the $R$ crystal field, we examine the case of no octahedral distortion. Therefore, we eliminate this term since it gives only a constant contribution. The effects of the O crystal field due to the actual TiO$_6$ distortion will be examined later. The second term $H_{s\tau}$ is the spin-pseudospin term. We introduce an additional term for the crystal field from the $R$ cations ($H_{\rm cry.}$). Finally, we obtain the Hamiltonian:
\begin{equation}
  H_{\rm eff} = H_{\rm cry.} + H_{s\tau}.
\end{equation}

It should be mentioned that our approach is appropriate for the systematic study on properties and mechanism of magnetic and orbital phase transitions since the origin of the stabilization of magnetic and orbital ordered states are attributed to the second-order perturbational energy gains with respect to the indirect $d$-$d$ transfers. In addition, their phase transitions are caused by the competition of these energy gains. The energy gains are easily evaluated in our approach by estimating the anisotropic electron-transfer amplitudes and the energy levels of the $3d$ orbitals.

\section{Results on LaTiO$_3$}

\begin{figure}[tdp]
\includegraphics[scale=0.4]{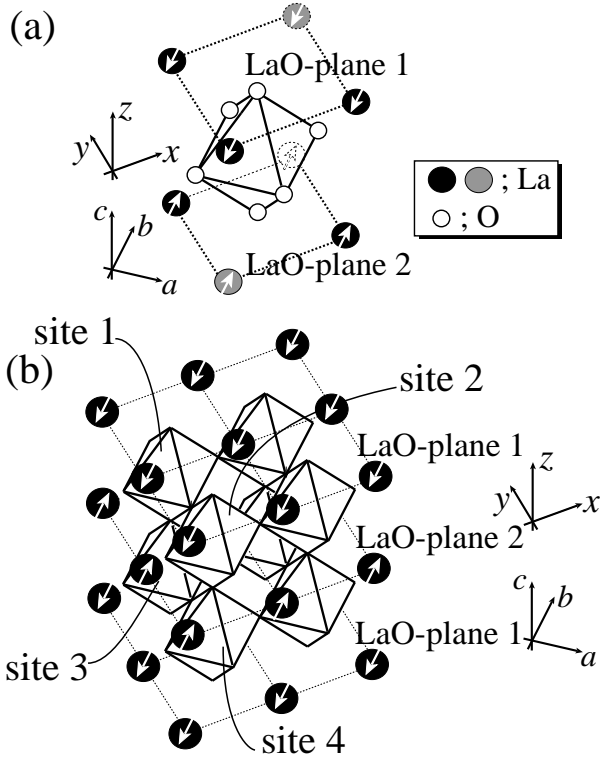}
\hfil
\caption{(a) Displacements of the La cations around a TiO$_6$ octahedron (site 1) in the GdFeO$_3$ structure are presented by arrows. In LaO-plane 1, they shift in the negative direction along the $b$ axis while in LaO-plane 2, they shift in the positive direction. The distances between the Ti ion and two La ions (gray circles) located in the $\pm(1,1,1)$-directions are decreased. As a result, the La ions generate a crystal field similar to the $D_{3d}$ crystal field with [1,1,1] trigonal axis (see text). Note that rotations of the octahedra due to the GdFeO$_3$-type distortion are not presented. (b) Stacking of the TiO$_6$ octahedra as well as shifts of the La ions are presented. Note that Ti and O ions are not explicitly presented in this figure.}
\label{reshift}
\end{figure}
In this section, we study effects of the crystal field due to the displacements of the La ions induced by the GdFeO$_3$-type distortion in LaTiO$_3$.
With the GdFeO$_3$-type distortion, the La ions shift mainly along the $(1,1,0)$-axis (namely, the $b$-axis), and slightly along the $(1,-1,0)$-axis (the $a$-axis).~\cite{MacLean79,Eitel86,Mizokawa99} There are two kinds of LaO-planes (plane 1 and plane 2) stacking alternatingly along the $c$-axis ($(0,0,1)$-axis) as shown in Fig.~\ref{reshift}. In plane 1, the La ions shift in the negative direction along the $b$-axis while they shift in the positive direction in plane 2. Consequently, the crystal field from the La ions is distorted from a cubic symmetry. For sites 1 and 2, the distances between the Ti ion and the La ions located in the $\pm(1,1,1)$-directions decrease while those along the $\pm(1,1,-1)$-directions increase (see Fig.~\ref{reshift} (a)). On the other hand, the distances along $\pm(1,1,-1)$-directions decrease while those along  $\pm(1,1,1)$-directions increase for sites 3 and 4. (In Fig.~\ref{reshift} (b), the stacking of each site is presented.) The changes of the other Ti-La distances are rather small. Since the nominal valence of the rare-earth ion is 3+, we expect that the Ti $3d$ orbitals directed along the shorter Ti-La bonds are lowered in energy because of the attractive Coulomb interaction. Without any distortions of the TiO$_6$ octahedra, the crystal field from the ligand oxygens has a cubic symmetry.

At this stage, when we introduce the crystal field from the La cations ($H_{R1}$) as a perturbation, the threefold degeneracy of the cubic-$t_{2g}$ levels splits. Here, for simplicity, we take $H_{R1}$ by assuming that there exist point charges with +3 valence on each La cation. The Coulomb interaction between an electron on a Ti $3d$ orbital and surrounding La$^{3+}$ ions is given by using a dielectric constant $\epsilon_{\rm TiLa}$ as follows,
\begin{equation}
 v(\vct {$r$})=-\sum_{i}
 \frac{Z_R e^2}{\epsilon_{\rm TiLa}|{\vct {$R$}}_i - {\vct {$r$}}|},
\end{equation}
where $\vct {$R$}_i$ expresses the coordinate of the $i$-th La ion, and $Z_R$(=+3) is the nominal valence of the La$^{3+}$ ion. We calculate the following matrix elements,
\begin{equation}
 \langle m|v(\vct {$r$})|m' \rangle= 
 \int {d\vct {$r$}} {{\varphi}^*}_{3d,m} v(\vct {$r$}){\varphi_{3d,m'}}
\end{equation}
with
\begin{eqnarray}
 \varphi_{3d,m}=R_{3d}({\vct r}) Y_{2m}(\theta,\phi),\\
 R_{3d}({\vct r})=\frac{4}{81\sqrt{30}}
                  \left( \frac{Z_M}{a_0} \right) ^{3/2}
                  {\rho}^2 {\rm e}^{-\frac{\rho}{3}}.
\end{eqnarray}
Here, the integer $m$(=$-2,...,2$) is the magnetic quantum number, $a_0$ is the Bohr radius, $\rho=\frac{Z_M}{a_0}r$, and $Z_M$(=+3) is the nominal valence of the Ti$^{3+}$ ion. The coordinates of the La ions are calculated by using the positional parameters and the cell constants obtained by the x-ray diffraction study.~\cite{MacLean79} After transforming the basis from $m$ to the $t_{2g}$ representations, we obtain the Hamiltonian $H_{R1}$ with $xy$-, $yz$- and $zx$-basis. These representations are defined in terms of the $x$-, $y$- and $z$-axes attached to each TiO$_6$ octahedron. On the basis of the analyses of thus obtained $H_{R1}$, we show in the following that the consideration of the La crystal field well explains the emergence of the AFM(G) ordering and the puzzling experimental findings in LaTiO$_3$.

\begin{figure}[tdp]
\includegraphics[scale=0.45]{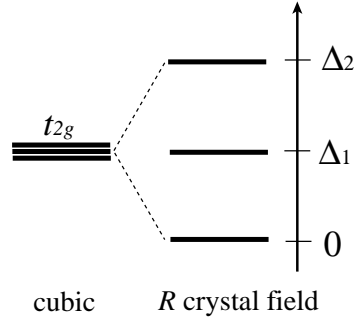}
\hfil
\caption{Energy-level structure in the crystal field due to the La cations. The threefold degenerate cubic-$t_{2g}$ levels split into three nondegenerate levels.}
\label{lvsplit}
\end{figure}
By diagonalizing $H_{R1}$, we can evaluate the $3d$-level energies and their representations. The threefold degenerate $t_{2g}$ orbitals split into three isolated levels as shown in Fig.~\ref{lvsplit}. The energy levels $\Delta_1$ and $\Delta_2$ are 0.77/$\epsilon_{\rm TiLa}$ eV and 1.58/$\epsilon_{\rm TiLa}$ eV, respectively. Here, it should be noted that since it is difficult to evaluate the local screening effect in a solid, it is hard to determine the value of effective dielectric constant $\epsilon_{\rm TiLa}$ without ambiguity. Therefore, in our calculation, $\epsilon_{\rm TiLa}$ is left as a variable.

The representations of the lowest levels at each site are specified by the linear combinations of the $xy$, $yz$ and $zx$ orbitals as follows,
\begin{eqnarray}
&{\rm site}&\quad 1; \quad a|xy\rangle+c|yz\rangle+b|zx\rangle, \nonumber \\
&{\rm site}&\quad 2; \quad a|xy\rangle+b|yz\rangle+c|zx\rangle, \nonumber \\
&{\rm site}&\quad 3; \quad a|xy\rangle-c|yz\rangle-b|zx\rangle, \nonumber \\
&{\rm site}&\quad 4; \quad a|xy\rangle-b|yz\rangle-c|zx\rangle.
\label{eqn:orbstrct1}
\end{eqnarray}
where $a^2$+$b^2$+$c^2$=1. Here, we note that there exists a mirror plane vertical to the $c$-axis, and the orbital structure has the same symmetry as the orthorhombic GdFeO$_3$-type distortion. 

\begin{figure}[tdp]
\includegraphics[scale=0.4]{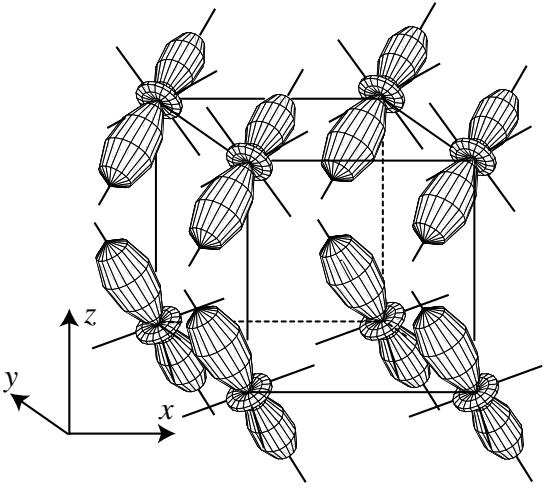}
\caption{Orbital structure in the $D_{3d}$ crystal fields with [1,1,1], [1,1,1], [1,1,$-$1] and [1,1,$-$1] trigonal axes at sites 1, 2, 3 and 4, respectively. The orbital wavefunctions at each site are $\frac{1}{\sqrt{3}}(xy+yz+zx)$, $\frac{1}{\sqrt{3}}(xy+yz+zx)$, $\frac{1}{\sqrt{3}}(xy-yz-zx)$ and $\frac{1}{\sqrt{3}}(xy-yz-zx)$, respectively. Note that tiltings due to the GdFeO$_3$-type distortion are not presented explicitly.}
\label{config3}
\end{figure}

For LaTiO$_3$, the coefficients take $a$=0.60, $b$=0.39 and $c$=0.69. The orbital wavefunctions at sites 1 and 2 are similar to $\frac{1}{\sqrt{3}}(xy+yz+zx)$, which is the lowest orbital for the $D_{3d}$ crystal field with $[1,1,1]$ trigonal axis. On the other hand, the orbital wavefunctions at sites 2 and 4 are similar to $\frac{1}{\sqrt{3}}(xy-yz-zx)$, which is the lowest orbital for the $D_{3d}$ crystal field with [1,1,$-$1] trigonal axis. Since there are four trigonal axes in an octahedron, there are several configurations of the trigonal axes at sites 1, 2, 3 and 4. The orbital structure realized in the crystal field of $H_{R1}$ is similar to that with the lowest orbitals in the $D_{3d}$ crystal fields with [1,1,1], [1,1,1], [1,1,$-$1] and [1,1,$-$1] trigonal axes at sites 1, 2, 3 and 4, respectively (see Fig.~\ref{config3}). In the previous study (Ref.~\citen{Mochizuki01b}), this trigonal-axes configuration was referred to as config. 3, and it was proposed that the orbital ordering in the $D_{3d}$ crystal fields with config. 3 strongly stabilizes the AFM(G) ordering, and well explains the puzzling experimental properties of LaTiO$_3$. In addition, recent dynamical mean-field calculations combined with first principles local density approximation (LDA+DMFT) show similar orbital orders with nondegenerate orbital ground state for LaTiO$_3$ and are consistent with the results obtained here.~\cite{Pavarini03,Craco03}

Here, we note that this orbital ordering likely induces the $D_{3d}$ distortions of the unit TiO$_6$ octahedra with the trigonal axes of config. 3. Recently, such distortions were actually observed by diffraction measurements.~\cite{Cwik03} One might argue that this TiO$_6$ distortion could occur by the usual Jahn-Teller mechanism. However, this is not true because the orbital degeneracy is already lifted by the $R$ crystal field and the Jahn-Teller mechanism is not effective if the GdFeO$_3$-type distortion (tilting of the TiO$_6$ octahedra) is present. The observed TiO$_6$ distortion is simply induced by the orbital ordering due to the $R$ crystal field. In this sense, the $t_{2g}$ degeneracy is lifted mainly by the $R$ crystal field, and the TiO$_6$ distortion occurs subsequently. Our approach in which the $R$ crystal field is taken first as the primary driving mechanism is based on the fact that the GdFeO$_3$-type distortion is stabilized from much more high-energy origin of the $R$-O covalency as is detailed in [2] of Sec. 5.

\begin{figure}[tdp]
\includegraphics[scale=0.45]{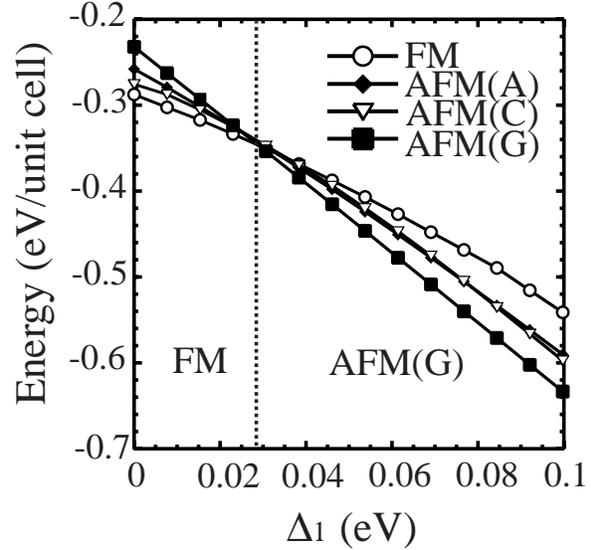}
\caption{Energies for several magnetic structures under the crystal field of $H_{R1}$ are plotted as functions of $\Delta_1$.}
\label{eneLaVMG}
\end{figure}

We next discuss the stability of the magnetic state in the crystal field of $H_{R1}$. For this purpose, we employ the effective spin-pseudospin Hamiltonian introduced in the previous section. Substituting $H_{R1}$ for the crystal field term $H_{\rm cry.}$, we calculate energies for several magnetic structures by applying the mean-field approximation. In Fig.~\ref{eneLaVMG}, we plot the calculated energies as functions of the magnitude of $\Delta_1$ (=$0.77/{\epsilon}_{\rm TiLa}$). The magnitude of $\Delta_1$ is varied because of the uncertainty due to ${\epsilon}_{\rm TiLa}$. In the region of $\Delta_1>$0.03 eV, the AFM(G) state is strongly stabilized relative to the other structures. 

In this region, the value of $\Delta_1$ is much larger than $k_{\rm B}T_{\rm N}$ so that the orbital occupation is restricted to the lowest orbitals irrespective of the spin structure. Now, we estimate the magnitude of the spin-exchange interaction for each Ti-Ti bond in the subspace of singly-occupied lowest orbitals. The spin-exchange constant $J$ for a Ti-Ti bond is represented as
$J=(E_{\uparrow\uparrow}-E_{\uparrow\downarrow})/2S^2$
with $E_{\uparrow\uparrow}$ and $E_{\uparrow\downarrow}$ being the mean-field energy gains for the Ti-Ti bond of $\uparrow\uparrow$- and $\uparrow\downarrow$-pairs, respectively. For LaTiO$_3$, the values of $J$ along the $x$-, $y$- and $z$-axes take as $J_x$=18.5 meV, $J_y$=18.5 meV and $J_z$=19.7 meV, respectively, which well reproduce the spin-wave spectrum with isotropic exchange constant of $\sim$15.5 meV. 

We have confirmed that the values of $J$ as well as the orbital wavefunctions obtained by the present calculation hardly change (within the accuracy of a few $\%$) even if the next nearest neighbor O and $R$ ions are taken into account, which suggests that the present result should be well established even with the ions further than the nearest neighbors since in reality, the ion-ion interaction is strongly screened as the distance increases. The robustness of the above results will be discussed in more detail in Sec. 5.

Here, we further note that the crystal field from the La cations has two origins. One is the Coulomb potential from charged La ions as we have studied above, and the other is the hybridization between the Ti $3d$ orbitals and unoccupied orbitals on the La ions. We next examine the latter effects for LaTiO$_3$. We construct the Hamiltonian for hybridization between the Ti $3d$ and La $5d$ orbitals ($H_{R2}$) by using the second-order perturbational expansion in terms of the transfers between the Ti $3d$ and La $5d$ orbitals ($t^{dd}$). The expression of the matrix element of $H_{R2}$ is
\begin{equation}
 \langle m|H_{R2}|m' \rangle= -
 \sum_{i,\gamma} \frac{t_{m;i\gamma}^{dd}t_{m';i\gamma}^{dd}}{\Delta_{5d}}.
\end{equation}
Here, the indices $m$ and $m'$ run over the cubic-$t_{2g}$ representations, $xy$, $yz$ and $zx$. The symbols $i$ and $\gamma$ are indices for the eight nearest-neighbor La ions and the fivefold La $5d$ orbitals, respectively. The symbol $\Delta_{5d}$ denotes the characteristic energy difference between the Ti $t_{2g}$ and La $5d$ orbitals. The transfer integral $t^{dd}$ is given in terms of Slater-Koster parameters $V_{dd{\sigma}}$, $V_{dd{\pi}}$ and $V_{dd{\delta}}$. It is assumed that these parameters are proportional to $d^{-5}$ with $d$ being the Ti-La bond length.~\cite{Harrison89} On the basis of the analyses of LDA band structure~\cite{Hamada02} and the photoemission spectra~\cite{Fujimori92}, we fix these parameters as $V_{dd{\sigma}}=-1.04$ eV, $V_{dd{\pi}}=0.56$ eV and $V_{dd{\delta}}=0$ eV for the Ti-La bond length of 3.5 $\AA$, and $\Delta_{5d}=3.6$ eV.

By diagonalizing thus obtained $H_{R2}$, we have evaluated the energy levels and their representations (see Table~\ref{tab:enelvl}). In the obtained energy level structure, the threefold $t_{2g}$ levels split into three nondegenerate levels, which is similar to that of $H_{R1}$. Since the representation of the lowest orbital of $H_{R2}$ is also similar to that of $H_{R1}$, the level splitting of $H_{R1}$+$H_{R2}$ is well expressed by the sum of these two contributions.~\cite{sec3R} Considering the fact that AFM(G) state is stabilized in the region of $\Delta_1>$0.03 eV, and $\Delta_1$ for $H_{R2}$ is $\sim$0.085 eV, the crystal field due to the hybridization between the Ti $3d$ and La $5d$ orbitals ($H_{R2}$) alone turns out to stabilize the AFM(G) spin structure strongly.

In addition, the value of $\Delta_1$ is sufficiently large as compared with the coupling constant of the LS interaction in Ti$^{3+}$ (${\zeta}_d$=0.018 eV).~\cite{Sugano70} Consequently, the crystal field from the La cations dominates over the LS interaction, resulting in the quenched orbital moment. Now, the experimentally observed reduction of the magnetic moment is attributed to the strong itinerant fluctuation near the metal-insulator phase boundary instead of the LS interaction as discussed in the literature~\cite{Mochizuki01b}. We will discuss this issue in $\S$ 5 in more detail.
\begin{table}[tdp]
\caption{Energy-level structures and representations
of the lowest orbitals for $H_{R1}$ and $H_{R2}$.}
\label{tab:enelvl}
\begin{tabular} {lcr}
      \hline
      \quad & $H_{R1}$         & \quad\quad\quad$H_{R2}$  \\
      \hline
      \hline
       $\Delta_1$ (eV) : & 0.77/$\epsilon_{\rm TiLa}$  & \quad\quad\quad0.08\\
       $\Delta_2$ (eV) : & 1.58/$\epsilon_{\rm TiLa}$  & \quad\quad\quad0.14\\
      \hline
       $a$ :   & 0.60   & \quad\quad\quad0.62 \\
       $b$ :   & 0.39   & \quad\quad\quad0.44 \\
       $c$ :   & 0.69   & \quad\quad\quad0.65 \\     \hline
\end{tabular}
\end{table}

\section{Results on $R$TiO$_3$ ($R$=La, Pr, Nd and Sm)}

In this section, we study the effects of the crystal field from the $R$ ions in $R$TiO$_3$ as functions of the GdFeO$_3$-type distortion (in other words, as functions of the size of $R$ ion). When the TiO$_6$ octahedra forming the perovskite lattice are distorted, the crystal field from the ligand oxygens is considered to be responsible for the electronic structure. However, with small or no atomic distortion of the octahedra, we expect that the crystal field from the $R$ ions sufficiently lifts the $t_{2g}$ and $e_g$ degeneracy of the $3d$ state, and plays a substantially important role on the stability of the magnetic and orbital states. The magnitude of the local distortion of the TiO$_6$ octahedron is expressed by the standard deviation from the mean for the three inequivalent Ti-O bond length.
\begin{figure}[tdp]
\includegraphics[scale=0.45]{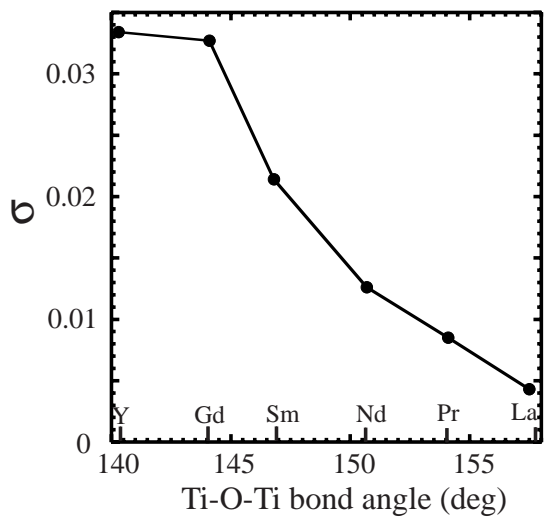}
\caption{Standard deviation from the mean for three inequivalent Ti-O bond length is plotted as a function of the Ti-O-Ti bond angle.~\cite{MacLean79}}
\label{stdevTiO}
\end{figure}
In Fig.~\ref{stdevTiO}, we plot the standard deviation ($\sigma$) obtained by the x-ray diffraction study~\cite{MacLean79} as a function of the Ti-O-Ti bond angle. The value for PrTiO$_3$ is obtained by the interpolation of LaTiO$_3$ and NdTiO$_3$. Focusing on the compounds with relatively small $\sigma$, we have studied relations between the GdFeO$_3$-type distortion and the $R$ crystal field systematically. We again derive the crystal field Hamiltonians based on the point charge model ($H_{R1}$) for $R$TiO$_3$ with $R$ being Pr, Nd and Sm by using the coordination parameters, and calculate the wavefunctions of the lowest orbitals. Here, the value of $\sigma$ for SmTiO$_3$ is rather large as compared with the other three compounds of $R$=La, Pr and Nd, which indicates relatively strong distortion of the TiO$_6$ octahedra. This means that the crystal field from the O ions in SmTiO$_3$ is deviated from the cubic symmetry, which would strongly affect the electronic structure in this compound. Therefore, we have to take effects of both Sm and O crystal fields into account. Thus, we first study the effects of the Sm crystal field ($H_{R1}$). Then, the effects of the distortion of the TiO$_6$ octahedra and the O crystal field will also be studied later.

\begin{figure}[tdp]
\includegraphics[scale=0.4]{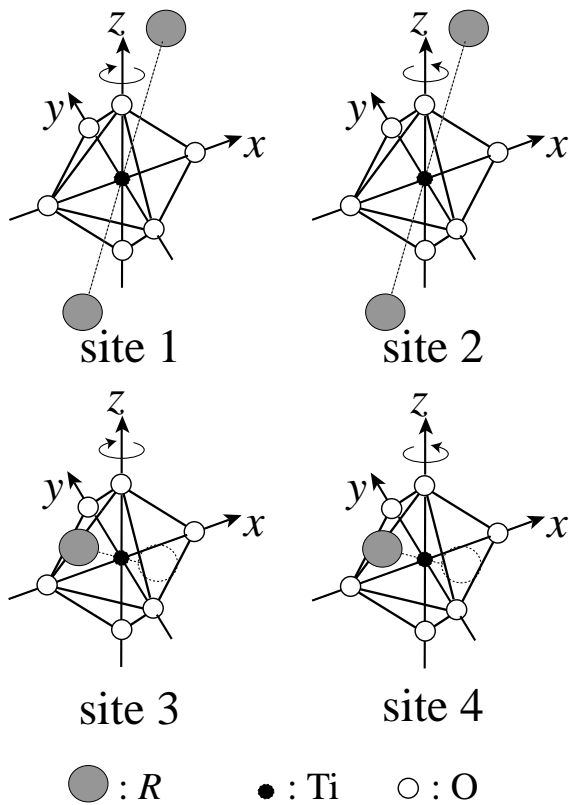}
\caption{Rotations of the TiO$_6$ octahedra around the $c$-axis in the GdFeO$_3$ structure are presented for sites 1-4. The closest two $R$ cations are also presented by gray and white large circles.}
\label{tilt}
\end{figure}
\begin{figure}[tdp]
\includegraphics[scale=0.45]{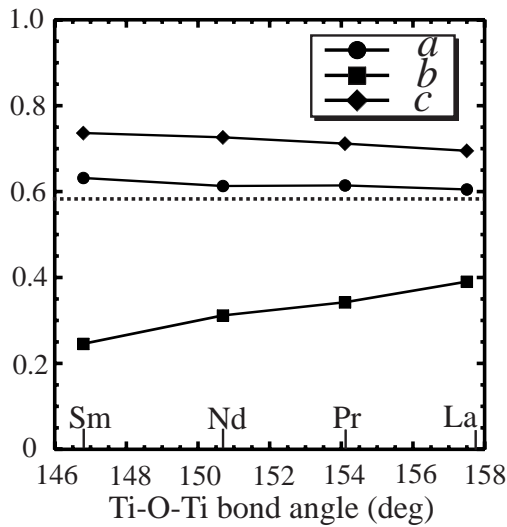}
\caption{Orbital wavefunctions at each Ti site are presented as functions of the Ti-O-Ti bond angle. The dashed line indicates 1/$\sqrt{3}$. When the bond angle decreases, the orbitals continuously approach the character of YTiO$_3$ given by Eq.~(\ref{eqn:orbstrct2}).}
\label{wavefncRCF}
\end{figure}
First, we study the orbital structures in the $R$ crystal fields. In LaTiO$_3$ with the smallest GdFeO$_3$-type distortion, the closest two $R$ ions are located nearly along the trigonal directions, and the orbital wavefunctions are approximately expressed by $\frac{1}{\sqrt{3}}(xy+yz+zx)$ and $\frac{1}{\sqrt{3}}(xy-yz-zx)$ as discussed in the previous section. The weights of the $xy$, $yz$ and $zx$ orbitals are almost the same in this compound. As the GdFeO$_3$-type distortion increases, the orbital wavefunctions should differ from those of LaTiO$_3$ since the locations of the $R$ ions are deviated from the trigonal directions. We expect that with increasing GdFeO$_3$-type distortion, the component of the $zx$ orbital gradually decreases at sites 1 and 3, while the $yz$-component decreases at sites 2 and 4. This expectation is based on the following discussion. Tilting of the TiO$_6$ octahedra in the GdFeO$_3$-type distortion is approximately expressed by rotations around the $b$- and $c$-axes (see also Fig.~\ref{tilt}). The rotation around the $c$-axis is clockwise at sites 1 and 3, and counterclockwise at sites 2 and 4. Therefore, at sites 1 and 3, distances between the $zx$-plane and the closest two $R$ cations increase with increasing distortion, while those for $yz$-plane increase at sites 2 and 4. As a result, the attractive potential between the $R$ cations and the $zx$ orbitals are reduced at sites 1 and 3, while those for the $yz$ orbitals are reduced at sites 2 and 4, resulting in the decrease of the corresponding orbital occupations.

In this discussion, only the rotations around the $c$-axis are taken into account. However, in the actual situation, the rotations around the $b$-axis also exist. Moreover, the locations of the $R$ ions are different from the ideal positions assumed in the above discussion. Thus, for more precise estimate, it is necessary to study the orbital structures by constructing the crystal field Hamiltonian $H_{R1}$ in which the experimentally measured position parameters of the $R$ and O ions are taken into account.

The lowest orbitals of $H_{R1}$ for these compounds are also expressed by Eq.~(\ref{eqn:orbstrct1}). In Fig.~\ref{wavefncRCF}, we show the coefficients $a$, $b$ and $c$ for the lowest orbitals of $H_{R1}$ at each Ti site as functions of the Ti-O-Ti bond angle. With relatively small GdFeO$_3$-type distortion, the orbital wavefunction has a form similar to that of the lowest orbital in the $D_{3d}$ crystal field, namely $\frac{1}{\sqrt{3}}(xy+yz+zx)$ or $\frac{1}{\sqrt{3}}(xy-yz-zx)$. As the GdFeO$_3$-type distortion increases, the component of the $zx$ orbital actually decreases at sites 1 and 3, while the $yz$-component decreases at sites 2 and 4, which is consistent with the above discussion.

The reductions of these orbital occupations indicate that with increasing GdFeO$_3$-type distortion, the crystal field from the $R$ cations tends to progressively stabilize the orbital structure in which sites 1, 2, 3  and 4 are occupied by the orbitals as follows:
\begin{eqnarray}
&{\rm site}&\quad 1; \quad
\frac{1}{\sqrt{2}}(|xy\rangle+|yz\rangle), \nonumber \\
&{\rm site}&\quad 2; \quad  
\frac{1}{\sqrt{2}}(|xy\rangle+|zx\rangle), \nonumber \\
&{\rm site}&\quad 3; \quad 
\frac{1}{\sqrt{2}}(|xy\rangle-|yz\rangle), \nonumber \\
&{\rm site}&\quad 4; \quad 
\frac{1}{\sqrt{2}}(|xy\rangle-|zx\rangle).
\label{eqn:orbstrct2}
\end{eqnarray}

In this orbital structure, there exists a mirror plane which is vertical to the $c$-axis. In fact, this orbital structure is the same as the orbital ordering which has been observed in YTiO$_3$ with large $d$-type Jahn-Teller distortion by several experiments such as NMR (Ref.~\citen{Itoh97,Itoh99}), polarized neutron scattering (Ref.~\citen{Ichikawa00,Akimitsu01}), and resonant x-ray scattering (Ref.~\citen{Nakao00}). Further, a recent resonant x-ray scattering study shows that the orbital state in SmTiO$_3$ has a twofold symmetry similarly to YTiO$_3$ and GdTiO$_3$, which is consistent with the present result.~\cite{Kubota00} It is interesting to note that, in our result, the character of the orbital symmetry in YTiO$_3$ is retained but quantitatively and continuously decreases toward the type of LaTiO$_3$. The detection of the orbital ordering in each compound predicted here can determine the validity of our theory.

\begin{figure}[tdp]
\includegraphics[scale=0.5]{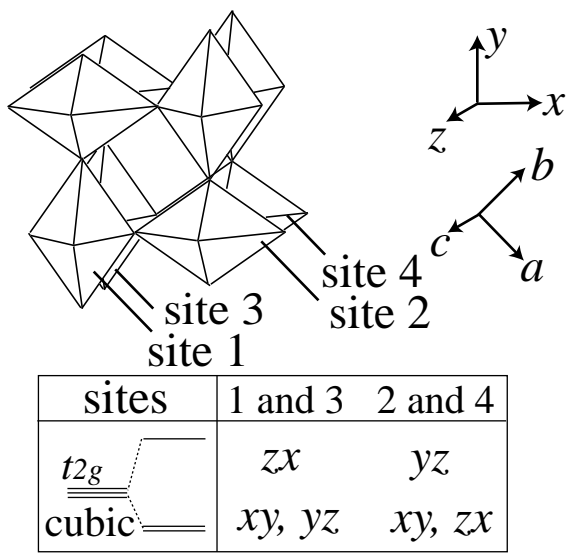}
\caption{$d$-type Jahn-Teller distortion. The threefold degenerate $t_{2g}$ levels split into twofold degenerate lower levels and a nondegenerate higher level at each Ti site. The ways of splitting at each site are also shown.}
\label{dJTdist}
\end{figure}
Now let us discuss the relations between the orbital state here obtained and the experimentally observed Jahn-Teller distortion in YTiO$_3$ and GdTiO$_3$. In the previous section, we have mentioned that in LaTiO$_3$, the orbital ordering stabilized due to the La crystal field favors the stacking of the trigonally distorted TiO$_6$ octahedra with [1,1,1] and [1,1,$-$1] trigonal axes alternatingly along the $c$-axis, which was actually detected experimentally. On the other hand, YTiO$_3$ and GdTiO$_3$ with strong GdFeO$_3$-type distortion show the $d$-type Jahn-Teller distortion in which the elongated axes of the octahedra are parallel along the $c$-axis as shown in Fig.~\ref{dJTdist}. As a result of this distortion, the $xy$ and $yz$ orbitals are lowered in energy at sites 1 and 3, while at sites 2 and 4, the $xy$ and $zx$ orbitals are lowered. In fact, the orbital wavefunctions in Eq.~(\ref{eqn:orbstrct2}) are represented by linear combinations of the twofold lower orbitals in the $d$-type Jahn-Teller distortion. This suggests that the $d$-type Jahn-Teller distortion realized in $R$TiO$_3$ ($R=$Y and Gd) is favored by the orbital ordering stabilized by the $R$ crystal field. In fact, a recent multiband $d$-$p$ model study has shown that with strong GdFeO$_3$-type distortion, the covalency between the O and $R$ ions induces the $d$-type Jahn-Teller distortion.~\cite{Mizokawa99} We stress that these two effects may cooperatively work to stabilize this distortion. It is interesting to point out that in the titanates, the $R$ ions in the GdFeO$_3$-type structure may control the octahedral distortions.

Here, we note that the orbital structure in SmTiO$_3$ may be slightly deviated from the orbitals presented here because of an additional effect of the O crystal field with relatively large distortion of the TiO$_6$ octahedra. More concretely, we expect that in SmTiO$_3$, the occupation of the $xy$ orbital is somewhat increased at all Ti sites as will be discussed later. 

\begin{figure}[tdp]
\includegraphics[scale=0.45]{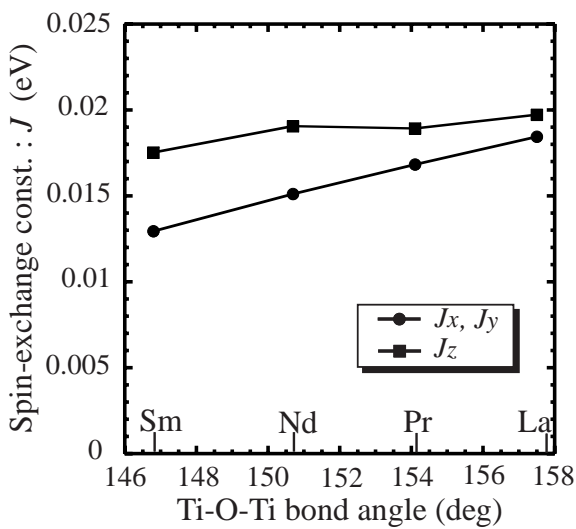}
\caption{Spin-exchange constants along the $x$-, $y$- and $z$-axes are plotted as functions of the Ti-O-Ti bond angle.}
\label{spinexJRCF}
\end{figure}

We also calculate the spin-exchange constant $J$ by assuming that the orbital occupation is restricted to the lowest orbital at every Ti site. In Fig.~\ref{spinexJRCF}, we plot the values of $J_x$, $J_y$ and $J_z$ as functions of the Ti-O-Ti bond angle. The AFM(G) spin structure is also reproduced in $R$TiO$_3$ with $R$ being Pr, Nd and Sm.
In LaTiO$_3$ with the smallest GdFeO$_3$-type distortion, a nearly isotropic spin coupling is realized. As the GdFeO$_3$-type distortion increases, the spin exchange gradually decreases. This is consistent with the gradual decrease of $T_{\rm N}$ as $R$ goes from La, Pr, Nd to Sm. In this calculation, the decrease of $J_x$ and $J_y$ is rather steep relative to that of $J_z$, resulting in the slightly anisotropic spin couplings. However, these subtle anisotropies can possibly be diminished or affected when we introduce the crystal field from the O ions as will be discussed later. It should be noted that these anisotropies will not necessarily be observed experimentally.

\begin{figure}[tdp]
\includegraphics[scale=0.45]{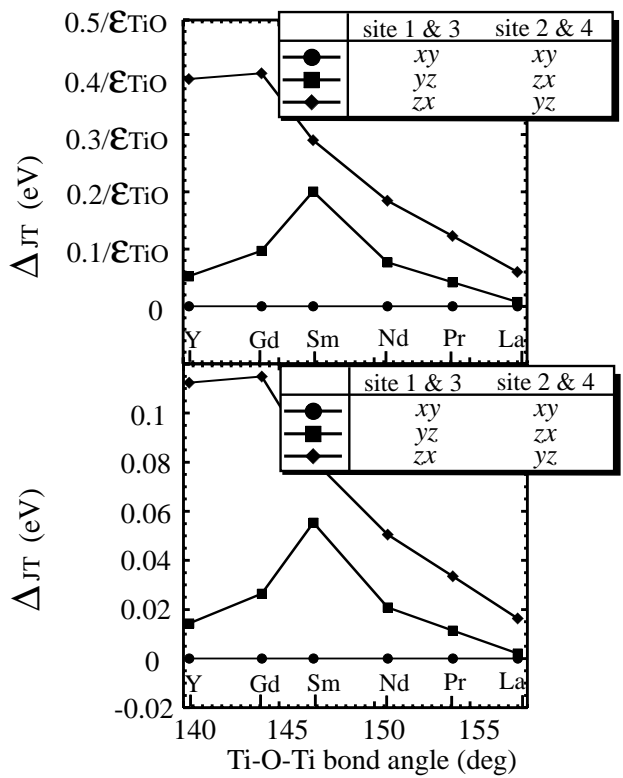}
\caption{Energy-level structures of the crystal field Hamiltonians $H_{L1}$ (upper panel) and $H_{L2}$ (lower panel) as functions of the Ti-O-Ti bond angle (see text).}
\label{levelJT}
\end{figure}
Here, we note that $T_{\rm N}$ for SmTiO$_3$ is about 50 K, which is decreased to 40$\%$ of $T_{\rm N}$ for LaTiO$_3$. However, the calculated spin-exchange constant is not so strongly depressed, and the value of $J_x$ for SmTiO$_3$ is about 70$\%$ of that for LaTiO$_3$. This is puzzling when we consider the fact that the model calculation based on the 3D Heisenberg model on the cubic lattice show that $T_{\rm N}$ is proportional to $J$.~\cite{Sandvik99}

In order to solve this problem, we study effects of the distortion of the TiO$_6$ octahedra. In the above analyses, we have examined the effects of the GdFeO$_3$-type distortion by assuming the undistorted TiO$_6$ octahedra as well as the cubic O crystal fields since the generic GdFeO$_3$-type distortion induces only tiltings and rotations of the octahedra, and does not distort them. However, in addition to the GdFeO$_3$-type distortion, distortions of the octahedra themselves are relatively strong in SmTiO$_3$ so that it is necessary to examine the effects of the O crystal field deviated from the cubic symmetry in order to clarify the electronic structure in this compound.

We derive the crystal field Hamiltonians for the repulsive Coulomb potential between an electron on the Ti $t_{2g}$ orbitals and the ligand O$^{2-}$ ions ($H_{L1}$) by assuming the form of the Coulomb interaction as
\begin{equation}
 v({\vct {$r$}})=\frac{2e^2}{{\epsilon_{\rm TiO}}
 |{\vct {$R$}}_i - {\vct {$r$}}|},
\end{equation}
where $\epsilon_{\rm TiO}$ is a dielectric constant, and ${\vct {$R$}}_i$ denotes the coordinates of the $i$-th O ions. 

In addition, the crystal field Hamiltonians for hybridization between the Ti $t_{2g}$ and O $2p$ orbitals ($H_{L2}$) is also derived. The expression of the matrix element of $H_{L2}$ is
\begin{equation}
 \langle m|H_{L2}|m' \rangle=
 \sum_{i,l} \frac{t_{m;il}^{pd}t_{m';il}^{pd}}{\Delta}.
\end{equation}
Here, the indices $m$ and $m'$ run over the cubic-$t_{2g}$ representations, $xy$, $yz$ and $zx$. The symbols $i$ and $l$ are indices for the six ligand O ions and the threefold O $2p$ orbitals, respectively. The symbol $\Delta$ stands for the charge-transfer energy. The transfer integral between the Ti $3d$ and O $2p$ orbitals ($t^{pd}$) is given in terms of Slater-Koster parameters $V_{pd{\sigma}}$, $V_{pd{\pi}}$. These parameters are calculated by assuming that they are proportional to $d^{-3.5}$ with $d$ being the Ti-O bond length.~\cite{Harrison89}

In Fig.~\ref{levelJT}, we plot relative energies of each level for $H_{L1}$ (upper panel) and those for $H_{L2}$ (lower panel). In the compounds with small GdFeO$_3$-type distortion, the differences of the energies for each level are rather small whereas in largely distorted GdTiO$_3$ and YTiO$_3$, one orbital is higher in energy relative to the other two orbitals. In GdTiO$_3$ and YTiO$_3$, the $xy$ and $yz$ orbitals are strongly lowered in energy at sites 1 and 3, and the $xy$ and $zx$ orbitals at sites 2 and 4 with the large $d$-type Jahn-Teller distortion. 

\begin{figure}[tdp]
\includegraphics[scale=0.5]{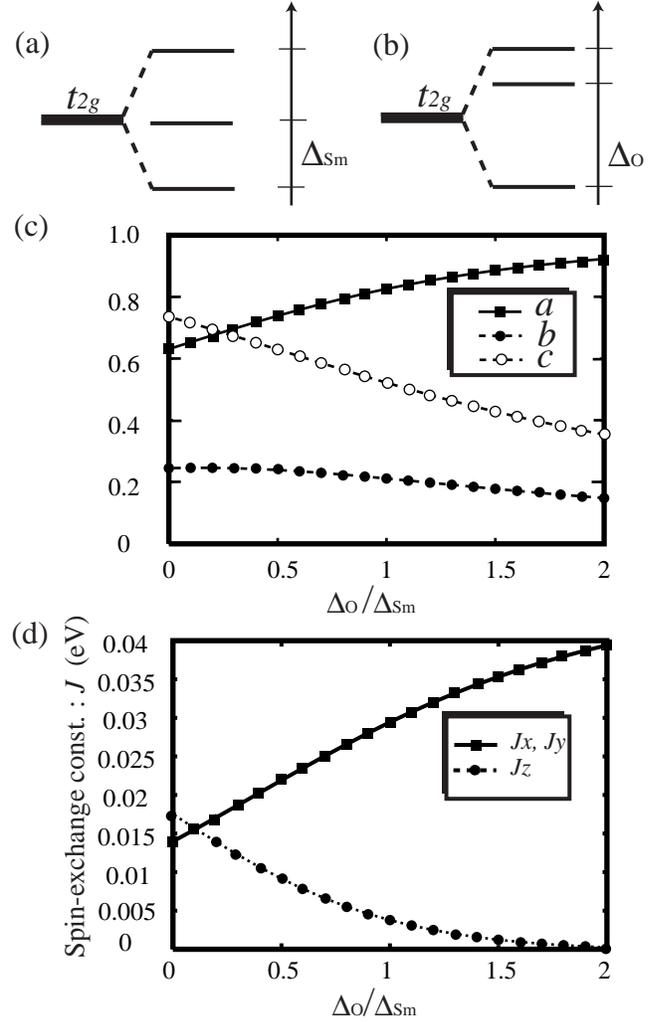}
\caption{(a) Energy-level scheme of the Sm crystal field Hamiltonian $H_{R1}$ for SmTiO$_3$. (b) Energy-level scheme of the O crystal field Hamiltonian $H_{L1}$ for SmTiO$_3$. (c) The lowest-orbital wavefunctions in the crystal fields from
both Sm and O ions ($H_{R1}$+$H_{L1}$) as functions of $\Delta_{\rm O}/\Delta_{\rm Sm}$. (d) Spin-exchange constants along the $x$-, $y$- and $z$-axes as functions of $\Delta_{\rm O}/\Delta_{\rm Sm}$.}
\label{SmOCRYST}
\end{figure}
In the case of moderately distorted SmTiO$_3$, the O crystal field lowers the $xy$ orbital at every Ti site, and both $yz$ and $zx$ orbitals are higher in energy. As a result, the distortions of the TiO$_6$ octahedra in SmTiO$_3$ favor the $xy$-occupation at every site. In this compound, the crystal field from the Sm ions and that from the O ions are strongly competing, and the low-energy orbital-spin structure is determined by this competition. In SmTiO$_3$, the Sm crystal field ($H_{R1}$) tends to lift the $t_{2g}$ degeneracy into three isolated levels as shown in Fig.~\ref{SmOCRYST} (a). The magnitude of splitting $\Delta_{\rm Sm}$ is 1.44/$\epsilon_{\rm TiSm}$ eV, and the lowest-orbital wavefunctions at each site are represented by Eq.~(\ref{eqn:orbstrct1}) with $a=$0.63, $b=0.25$ and $c=0.74$. On the other hand, if only the O crystal field ($H_{L1}$) exists, the $t_{2g}$ degeneracy splits into lower $xy$ orbital and higher $yz$ and $zx$ orbitals with $\Delta_{\rm O}$ of 0.20/$\epsilon_{\rm TiO}$ eV as shown in Fig.~\ref{SmOCRYST} (b). Because of the uncertainty due to $\epsilon_{\rm TiSm}$ and $\epsilon_{\rm TiO}$, it is difficult to estimate the strength of these two crystal fields qualitatively. Then, we study the orbital structure and the spin coupling by varying the ratio of the strength of these two crystal fields (in other words, by varying $\Delta_{\rm O}/\Delta_{\rm Sm}$). The lowest-orbital wavefunctions are also expressed by Eq.~(\ref{eqn:orbstrct1}) even when both O and Sm crystal fields ($H_{R1}$ and $H_{L1}$) exist. In Fig.~\ref{SmOCRYST} (c), we plot the coefficients $a$, $b$ and $c$ as functions of $\Delta_{\rm O}/\Delta_{\rm Sm}$. This figure shows that the $xy$-orbital occupation increases with increasing $\Delta_{\rm O}/\Delta_{\rm Sm}$, namely, with increasing strength of the O crystal field.

The increased $xy$-occupation is expected to cause a two-dimensional anisotropy in the spin coupling in which a strong AFM coupling is realized in the $ab$-plane while along the $c$-axis, the AFM coupling is rather weak. In Fig.~\ref{SmOCRYST} (d), the spin-exchange constants $J_x$, $J_y$ and $J_z$ are plotted as functions of $\Delta_{\rm O}/\Delta_{\rm Sm}$. In this figure, the value of $J_z$ rapidly decreases while $J_x$ and $J_y$ increase as $\Delta_{\rm O}/\Delta_{\rm Sm}$ increases. In particular, in the region of $\Delta_{\rm O}/\Delta_{\rm Sm} \ge 1$, the value of $J_z$ becomes almost zero while $J_x$ and $J_y$ take relatively large value, and the ratio $J_z/J_x$ becomes smaller than 0.13. This means that when the strength of the O crystal field is comparable to, or is larger than the Sm crystal field, an almost perfect two-dimensional spin coupling is realized. These facts indicate that the increased $xy$-occupation due to the O crystal field actually causes the two-dimensional anisotropy in the spin coupling. The observed depression of $T_{\rm N}$ in SmTiO$_3$ may be attributed to this two-dimensional anisotropy. We predict that this two-dimensional anisotropy may be observed in SmTiO$_3$.

In addition, the crystal field from the Sm ions and that from the O ions make opposite contributions to the anisotropy of the spin coupling. As shown in Fig.~\ref{SmOCRYST}, without the O crystal field (namely, $\Delta_{\rm O}/\Delta_{\rm Sm}$=0), the Sm crystal field favors the anisotropy in which spin exchange along the $c$-axis is slightly larger than that in the $ab$-plane. On the other hand, the O crystal field favors the in-plane anisotropy. Figure~\ref{SmOCRYST} shows that the two-dimensional anisotropy due to the O crystal field dominates over the small contribution to the anisotropy from the $R$ crystal field even in the region of small $\Delta_{\rm O}/\Delta_{\rm Sm}$.

In SmTiO$_3$, we expect that both crystal field from the Sm ions and that from the O ions play important roles on the electronic state, and the orbital-spin structure in the Sm crystal field is strongly affected by the O crystal field because of the relatively strong TiO$_6$ distortion. It is essentially important to study the orbital state by quantitative estimate of the crystal field from both O and Sm ions to elucidate the orbital-spin structure of SmTiO$_3$, which is left for further study.

In the compounds of $R=$Pr and Nd, if we neglect the effects of the octahedral distortion, the calculated spin-exchange constants show very small anisotropies in which the AFM spin coupling along the $c$-axis is slightly larger than that in the  $ab$-plane as shown in Fig.~\ref{spinexJRCF}. However, even when the magnitude of the level splitting is considerably small, the O crystal field in PrTiO$_3$ and that in NdTiO$_3$ also favor the $xy$-occupation, which generates the AFM coupling in the $ab$-plane. Under these circumstances, the slight anisotropy favored in the $R$ crystal field should be diminished or weakened by the subsequent and opposite contribution from the O crystal field. Thus, we expect that the AFM coupling in PrTiO$_3$ and that in NdTiO$_3$ can be rather isotropic. The quantitative estimates of the competition between $R$ and O crystal fields in these compounds are left for further studies.

We finally mention that when the crystal field from the O ions due to the TiO$_6$ distortion is superimposed, the dominance of the $R$ crystal field is taken over by the O crystal field at $R$=Sm. Through severe competition at $R$=Sm, the O crystal field deviated from the cubic symmetry due to the $d$-type Jahn-Teller distortion comes to control the orbital-spin structure at $R$=Gd and Y. We note that in the present analyses, we have focused on the effects of the $R$ crystal fields on the cubic-$t_{2g}$ degeneracy, and the Jahn-Teller distortion is not taken into account so that the orbital-spin structure in the FM region with strong $d$-type Jahn-Teller distortion has not been reproduced. Therefore, a recent puzzling finding of isotropic ferromagnetism in YTiO$_3$~\cite{Ulrich02} is not resolved in this study and is left for further study.

\section{Discussion}

In this section, we will discuss five points on our proposal. Before the detailed discussion, we first summarize these points in the following:\\

\noindent
[1]: Our theory succeeded for the first time in explaining puzzling experimental results in a unified way. The available experimental results will be discussed in the light of our theory.

\noindent
[2]: A universal mechanism for controlling electronic structure of the perovskite-type transition-metal oxides, which has been overlooked is proposed. The mechanism here proposed is substantially different from the usual Jahn-Teller mechanism.

\noindent
[3]: We will also discuss the accuracy of the numerical results and the reliability of our predictions about orbital structures as well as the spin-coupling isotropy.

\noindent
[4]: Based on the present results together with the previous ones,~\cite{Mochizuki00,Mochizuki01a,Mochizuki02,Mochizuki01b,Mochizuki03} the overall features of the magnetic phase diagram for the titanates will be discussed.

\noindent
[5]: The observed magnetic moment of 0.45 $\mu_{\rm B}$ in LaTiO$_3$~\cite{Goral82} has been sometimes considered to be small by comparing with the value of $\sim$0.85 $\mu_{\rm B}$ in the spin-wave theory on the 3D Heisenberg model. We will discuss the origin of this reduction.\\

\noindent
$\bf [1]$: In the present paper, we have shown that a very simple mechanism of the $R$ crystal field here, we proposed and never been considered seriously before, accounts for available experimental results with no contradiction. Since the perovskite titanates attract an enormous amount of interest, a number of experiments have been done so far. However, there has been no theory which can explain these experimental results in a unified way. As discussed in the present paper, the following experimental results are explained within our theory for the first time.\\

\noindent
(1): Emergence of the AFM(G) state\\
The emergence of AFM(G) state in LaTiO$_3$ in which the $t_{2g}$ orbitals have been considered to be degenerate, has long been puzzling.
We have shown in this paper that the AFM(G) state is realized with quenched orbital degrees of freedom due to the crystal field from the La ions. Further, the analyses of the $R$ crystal fields in $R$TiO$_3$ with $R=$Pr, Nd and Sm have also reproduced the AFM(G) ordering.\\ 

\noindent
(2): Neutron scattering\\
Recent neutron scattering experiment showed the spin-wave spectrum characterized by isotropic spin coupling with $J$ of 15.5 meV as well as a small spin gap for LaTiO$_3$.~\cite{Keimer00} These characteristics are well reproduced in our picture. Since the lowest orbital state in the La crystal field is approximately represented by the linear combination of the $xy$, $yz$ and $zx$ orbitals with the same weights, nearly isotropic electron transfers along $x$, $y$ and $z$ directions are realized. The isotropic spin coupling is caused by this isotropy of transfer amplitudes. In addition, we have shown that the crystal field splitting is much larger than the coupling constant of the LS interaction, and the resultant quenched orbital moment leads to the observed small spin gap.\\ 

\noindent
(3): Behavior of $T_{\rm N}$ in the magnetic phase diagram\\ 
According to the experimentally obtained magnetic phase diagram,~\cite{Greedan85,Katsufuji97,Goral82,Okimoto95} $T_{\rm N}$ monotonically decreases as $R$ goes from La, Pr, Nd to Sm with increasing GdFeO$_3$-type distortion. The calculation of the spin coupling constant $J$ based on our theory has well reproduced this behavior.\\ 

\noindent
(4): Resonant x-ray scattering\\
Recent resonant x-ray scattering experiment showed the evidence of an orbital ordering in the AFM(G) compounds of $R$TiO$_3$ (R=La, Pr, Nd and Sm), and that the symmetry of the orbital ordering is the same as that in the FM compounds YTiO$_3$ and GdTiO$_3$.~\cite{Kubota00} Our calculation of the orbital states based on our theory well accounts for this result. \\

\noindent
(5): X-ray diffraction measurement\\
We have shown that in LaTiO$_3$, the crystal fields from the La cations stabilize an orbital ordering in which $\frac{1}{\sqrt{3}}(xy+yz+zx)$, $\frac{1}{\sqrt{3}}(xy+yz+zx)$, $\frac{1}{\sqrt{3}}(xy-yz-zx)$ and $\frac{1}{\sqrt{3}}(xy-yz-zx)$ are approximately occupied at four Ti sites in the orthorhombic unit cell in an alternating fashion along the $c$-axis. This orbital ordering is expected to induce nearly trigonal distortions of the TiO$_6$ octahedra with trigonal axes of [1,1,1], [1,1,1], [1,1,$-$1] and [1,1,$-$1] at each Ti site. These distortions were actually observed by recent x-ray diffraction measurements.~\cite{Cwik03} In addition, we have pointed out that the $d$-type Jahn-Teller distortions in YTiO$_3$ and GdTiO$_3$ is also favored by the orbital orderings expected in the $R$ crystal field.\\

\noindent
(6): NMR\\
A NMR spectrum for LaTiO$_3$ proved to be well reproduced if we take the orbital ordering model predicted in our theory.~\cite{Kiyama03}\\ 

We note that there exists no other theory which explains these experimental results in such consistent ways, and that our theory succeeded in the explanation for the first time.\\

\noindent
$\bf [2]$: We also emphasize the importance of our proposal as a universal mechanism determining the electronic structure in the perovskite-type transition-metal oxides. Through the present study, we have pointed out for the first time that in the perovskite-type transition-metal oxides, a crystal field from the $R$ ions caused by the GdFeO$_3$-type structure substantially lifts the 3$d$ orbital degeneracy. This mechanism of lifting the orbital degeneracy competes with the usual Jahn-Teller mechanism, and eventually plays an important role on controlling the orbital-spin structures. 

Our proposal includes two important points. One point is the significance of the $R$ ions. The perovskite-type transition-metal oxides have often been studied by employing some models which include effects of the strong electron correlations such as the Hubbard model, $d$-$p$ model, $t$-$J$ model and so on by assuming only the crystal fields on 3$d$ orbitals from the surrounding octahedron. In these models, effects of the crystal field due to the $R$ ions have been usually overlooked. However, we have shown that it is important to consider the roles of the $R$ ions when we study the electronic structure of these compounds. 

Another important point is a new role of the GdFeO$_3$-type distortion. In the GdFeO$_3$-type distortion, TiO$_6$ octahedra tilt alternatingly, which decreases the $M$-O-$M$ bond angle from 180$^{\circ}$. This distortion has been focused from the aspect of the bandwidth control since the decrease of the $M$-O-$M$ angle significantly reduces transfer integrals between neighboring transition-metal ions mediated by the oxygen ions. On the other hand, an octahedron itself does not atomically distort in the generic GdFeO$_3$-type distortion so that the O crystal field maintains a cubic symmetry. Therefore, it had been considered that the distortion does not lift the $t_{2g}$ and $e_g$ degeneracies. However, our result indicates that in reality, the distortion necessarily lifts the degeneracy through generating the $R$ crystal field, and determines the orbital-spin state. This direct effect on the electronic structure has been overlooked so far. 

Here, it should be noted that this mechanism is not a usual Jahn-Teller one. In the Jahn-Teller systems, 3$d$ electrons on the transition-metal ions and the lattice degree of freedom strongly couple to each other, and the lattice spontaneously distorts to lift the 3$d$ degeneracy and lower the electronic energy. On the other hand, the GdFeO$_3$-type distortion is caused by interactions between the $R$ ions and O ions. More concretely, Woodward showed that the octahedral tilting is induced by an energy gain in $R$-O covalency bonding in the perovskite compounds with $R$-site mismatching, and subsequent $R$-site shifts occur by $R$-O ionic interactions.~\cite{Woodward97} In Ref.~\citen{Woodward97}, it has been shown that the displacement of the $R$ ions is a direct consequence of the GdFeO$_3$-type tilting through $R$-O coupling.  We note that this distortion exists irrespective of the 3$d$ electronic state, and has nothing to do with any spontaneous lift of the electronic degeneracy in contrast with the Jahn-Teller mechanism. The lifting of the degeneracy of the Ti 3$d$ orbitals occurs just as a consequence of the crystal field from the $R$ ions. In this sense, the mechanism we have proposed is not a Jahn-Teller but an overlooked one, which is inherent and universal with GdFeO$_3$-type distortion.\\

\noindent
$\bf [3]$: In the present study, we have examined the point charge model with the nearest neighbor 8 $R$ ions in the case of degenerate $t_{2g}$ orbitals in the cubic crystal field from the nearest neighbor 6 O ions. The calculation of the spin-exchange constant well reproduced the experimentally measured isotropic superexchange parameters in LaTiO$_3$. Now, let us discuss the adequacy of the approximation used and the robustness of the predicted isotropy.

First, we would like to stress that the present results would not be changed even if further ions would be taken into account. Our approximate treatment with the nearest neighbor ions actually provides correct results not only qualitatively but also quantitatively. 

In order to confirm this, we have calculated the $t_{2g}$ splittings, the orbital wavefunctions and the spin-exchange constants by using a point charge model with additional next-nearest-neighbor 24 $R$ and 24 O ions. The differences between thus obtained values and those obtained by the point charge model with only nearest neighbor ions are considerably small (within a few $\%$) as shown in the following:\\

\noindent
With only nearest neighbor O and $R$ ions:
\begin{itemize}
\item $t_{2g}$ splitting ($\Delta_1$): 0.77/$\epsilon_{\rm LaTi}$ 
\item wavefunction: 0.605$xy$ + 0.390$yz$ + 0.695$zx$ 
\item spin exchange: $J_{ab}=$18.5 meV, $J_c=$19.7 meV, 
\end{itemize}

\noindent
With additional next nearest neighbor O and $R$ ions:
\begin{itemize}
\item $t_{2g}$ splitting ($\Delta_1$): 0.75/$\epsilon_{\rm LaTi}$ 
\item wavefunction: 0.608$xy$ + 0.396$yz$ + 0.688$zx$ 
\item spin exchange: $J_{ab}=$18.9 meV, $J_c=$19.4 meV. 
\end{itemize}

These results are calculated by assuming the same dielectric constant $\epsilon_{\rm TiLa}$ for both Coulomb repulsion from the nearest neighbor ions and that from the next nearest neighbors. In this sense, strong screening effects in a solid are neglected so that the differences are overestimated. Since the screening effects strongly depress the Coulomb interaction as the ion-ion distance increases, we expect that the difference should be much smaller if we take the screening properly into account.

Here, we note that the purpose of the present paper is not a quantitatively precise estimate of the potential from the surrounding ions, but the proposal of a qualitative and transparent picture for the novel mechanism in which $R$ cations lift the $t_{2g}$ degeneracy and determine the orbital-spin structure in the perovskite titanates. Nevertheless, it should be emphasized that the point charge model with the nearest neighbor $R$ and O ions here we employed gives sufficiently correct results also quantitatively. 

In addition, in the present paper we have shown that the hybridization between the La $5d$ and Ti $3d$ orbitals works cooperatively with the Coulomb potential from the nearest neighbor $R$ ions on the stabilization of the above orbital ordering and the AFM(G) state, and the hybridization alone has proved to strongly stabilize the orbital-spin structure. This fact also supports that the consideration of the Coulomb potential from further $R$ ions hardly changes the results. 

We also note that the predicted isotropy is robust even when we consider the slight TiO$_6$ distortion, which is actually realized in LaTiO$_3$. In our analyses, the TiO$_6$ octahedron is assumed to be undistorted. We consider that the actual orbital state may be slightly deviated from that calculated here because of the small TiO$_6$ distortion. Actually, in the paper by Cwik $et$ $al$., by considering the small octahedral distortion observed in their experiment, the orbital state is calculated as $0.77(\frac{1}{\sqrt{2}} (zx+yz)) + 0.636xy = 0.636xy+0.54yz+0.54zx$. This is also similar to the $\frac{1}{\sqrt{3}}(xy+yz+zx)$ orbital, but is slightly different from our result. However, in the previous study,~\cite{Mochizuki01b} we have evaluated the spin-exchange constant when the perfect $\frac{1}{\sqrt{3}}(xy+yz+zx)$ orbital is occupied, and we have shown that the spin coupling is also almost isotropic and the anisotropy ($J_{ab}/J_c$) is less than 10 $\%$. This result together with the present one suggests that the isotropy is hardly affected by the slight deviation of the orbital state from $\frac{1}{\sqrt{3}}(xy+yz+zx)$. Therefore, we conclude that the isotropy predicted in our study is robust.\\

\begin{figure}[tdp]
\includegraphics[scale=0.4]{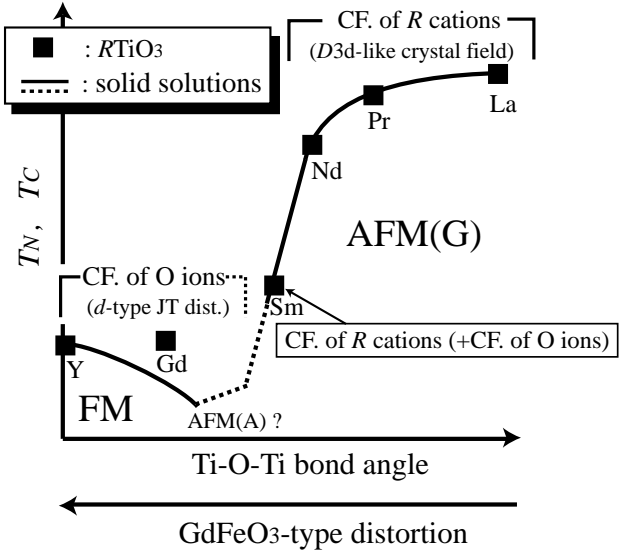}
\caption{Schematic magnetic phase diagram for $R$TiO$_3$ and solid-solution systems La$_{1-x}$Y$_x$TiO$_3$ in the plane of temperature and the GdFeO$_3$-type distortion. At $R=$La, Pr and Nd with relatively small GdFeO$_3$-type distortions, the crystal field (CF) from $R$ cations dominantly determines the electronic structures, while in significantly distorted GdTiO$_3$ and YTiO$_3$, the O crystal field dominates. In SmTiO$_3$, both crystal fields strongly compete.}
\label{magphase2}
\end{figure}

\noindent
$\bf [4]$: We next discuss the experimentally obtained magnetic phase diagram~\cite{Greedan85,Katsufuji97,Goral82,Okimoto95} based on the present results together with the previous ones~\cite{Mochizuki00,Mochizuki01a,Mochizuki02,Mochizuki01b,Mochizuki03} (see also Fig.~\ref{magphase2}). In LaTiO$_3$ with the smallest GdFeO$_3$-type distortion, since the atomic distortion of the TiO$_6$ octahedra is rather small, the crystal field from the La ions dominantly determines the Ti $3d$ state. As a result, LaTiO$_3$ exhibits a AFM(G) ordering with the lowest orbital occupation in the nearly-trigonal La crystal field. The AFM(G) ordering due to the $R$ crystal field is also realized in the compounds of Pr, Nd and Sm with relatively small octahedral distortion. In addition, as $R$ goes from La, Pr to Nd, $T_{\rm N}$ gradually decreases with decreasing spin-exchange constant. In SmTiO$_3$, the TiO$_6$ octahedra are moderately distorted so that the competition between the crystal field from the O ions and that from the Sm ions comes about. In SmTiO$_3$, the O crystal field works as increasing the $xy$-orbital occupation in all Ti sites. As a result, the AFM(G) spin coupling becomes to have a two-dimensional anisotropy. The depression of $T_{\rm N}$ of SmTiO$_3$ may be attributed to this anisotropy. 

In YTiO$_3$ and GdTiO$_3$ with large GdFeO$_3$-type distortion, the electronic structure is dominantly determined by the crystal field from the O ions with the largely distorted TiO$_6$ octahedra. These compounds show a FM ground state with an orbital ordering in the strong $d$-type Jahn-Teller distortion. In the previous studies,~\cite{Mochizuki00,Mochizuki01a} we have shown that the FM coupling in the vicinity of the AFM-FM phase boundary has a strong two-dimensional character, resulting in the depressed $T_{\rm C}$. The puzzling second-order like behavior of the AFM-FM transition can be understood from these depressions of $T_{\rm N}$ and $T_{\rm C}$. In the previous studies,~\cite{Mochizuki00,Mochizuki01a} we have also pointed out that although there is no stoichiometric system between SmTiO$_3$ (AFM(G), Ref.~\citen{Amow98}) and GdTiO$_3$ (FM), the AFM(A) state might be observed in the solid-solution systems such as La$_{1-x}$Y$_x$TiO$_3$ and Gd$_{1-x}$Sm$_x$TiO$_3$ sandwiched by the AFM(G) and FM phases.\\

\noindent
$\bf [5]$: We finally discuss the experimentally observed reduction of the ordered moment in LaTiO$_3$. The value of the ordered moment is about 0.45 $\mu_{\rm B}$, which is reduced from spin-only moment of 1 $\mu_{\rm B}$. Since a neutron scattering study showed that the LS interaction is irrelevant to this reduction, the origin has been a puzzling issue. Considering that LaTiO$_3$ with a small insulating gap of $\sim$0.2 eV is located near the metal-insulator phase boundary,~\cite{Okimoto95,Katsufuji95} we have attributed this reduction to the strong itinerant fluctuations. Within our effective Hamiltonian, the charge fluctuations are completely neglected in the limit of the strong Coulomb repulsion. In addition, the quantum fluctuations are also neglected in the mean-field approximation which we have used. Here, let us discuss this assignment by considering available numerical and analytical results on the Hubbard models.

The Hubbard model is characterized by the ratio $U/t$. The value of $U/t$ for LaTiO$_3$ can be estimated as follows: The characteristic energy of the nearest neighbor Ti $t_{2g}$ transfer $t$ is expressed as $V_{pd\pi}^2/\Delta$ based on the second-order expansion with respect to the transfer integrals between the 
Ti $t_{2g}$ and O $2p$ orbitals. The value is estimated as 0.33 eV by using the parameter values used in our calculation, where $\Delta$=5.5 eV and $V_{pd\pi}$=1.33 eV. Therefore, the ratio $U/t$ takes approximately 12 for the multiplet-averaged Coulomb interaction $U$ of 4 eV.  
  
The equal time magnetic structure factor, which has the form; 
\begin{equation}
S(\vct{$Q$})=\frac{1}{N} \sum_{i,j}e^{i\vct{$Q$}({\vct{$R$}}_i-{\vct{$R$}}_j)}
\langle (n_{i\uparrow}-n_{i\downarrow})
(n_{j\uparrow}-n_{j\downarrow}) \rangle,\\
\end{equation}
with $\vct{$Q$}$ being an AF wave vector, is related to the AF ordered moment $m$ through the formula: $\frac{S(\vct{$Q$})}{N} = m^2 + f(N)$. This quantity for the 3D Hubbard model has been calculated in several QMC simulations in small-sized cubic lattices by using the grand canonical method.~\cite{Staudt00,Hirsch87} However, because of the limitation of the cluster size, there exists considerable ambiguity in the extrapolation to the thermodynamic limit so that a reliable estimate of $S(\vct{$Q$})$ and that of the ordered moment are difficult in available calculations.

At this stage, we discuss these quantities based on some approximate results on the 3D Hubbard model or numerical results on other models. Then, we present some possible mechanisms of the reduction of the ordered moment. In particular, we discuss by focusing on effects of (1) charge fluctuations, (2) orbital degrees of freedom, and (3) imperfect nesting of the Fermi surface. 

We first consider effects of the charge fluctuations. An approach based on the Feynman's variational principle has been applied to the 3D Hubbard model.~\cite{Kakehashi86} This study shows that at $U/t$=12, a relatively large charge fluctuation remains, which is approximately 40$\%$ of that at $U/t$=0. Owing to this charge fluctuation, the ordered moment at $U/t$=12 is reduced to about 80$\%$ of the full moment. Here, we note that since this approach neglects quantum fluctuations, the ordered moment is saturated to 1 $\mu_{\rm B}$ in the limit of infinite $U$ (namely, in the limit of vanishing charge fluctuation). In the spin-wave theory on the 3D Heisenberg model, the ordered moment in the $U\to\infty$ limit diminishes to $\sim$0.85 $\mu_{\rm B}$ owing to the quantum fluctuations. Therefore, we naively expect that the ordered moment can diminish to $\sim$0.7 $\mu_{\rm B}$ when both quantum and charge fluctuations are properly taken into account. Although the charge fluctuations significantly reduce the ordered moment, this value is still a little large relative to the observed ordered moment of 0.45 $\mu_{\rm B}$, indicating a necessity of additional mechanisms.  

The orbital degrees of freedom can further reduce the ordered moment. In the orbitally degenerate systems, the AFM spin-exchange interaction is considerably reduced as compared with a single band case because of the FM contribution from electron transfers between neighboring occupied and unoccupied orbitals and the Hund's rule coupling. Therefore, the spin correlation in such systems is rather weak relative to the nondegenerate systems. The resultant relative increase of the charge fluctuations may play a role on the reduction of the ordered moment. Actually, the reduction in the orbitally degenerate system has been studied by Ole\'s.~\cite{Oles83} In this work, the Hubbard model with fivefold degenerate orbitals and isotropic transfers is studied by a Gutzwiller-type variational approach, and the ordered moment is calculated as a function of ($U$+$6j$)/$W$ for various values of $j/U$ where $j$ and $W$ are the Coulomb exchange interaction and the bandwidth, respectively. If we assume the band width of the 3D single-band Hubbard model ($W$=12$t$), the values of ($U$+$6j$)/$W$ and $j/U$ for LaTiO$_3$ are estimated as $\sim$1.75 and $\sim$0.1, respectively, where $U$=4 eV, $j$=0.46 eV and $t$=0.33 eV. The result shows that for these parameters, the ordered moment takes a considerably small value of $\sim$0.45-0.5 $\mu_{\rm B}$. In LaTiO$_3$, the crystal field from the $R$ cations works as lifting the degeneracy of the $t_{2g}$ orbitals at Ti sites. In this sense, this compound may interpolate between a single-band limit and a completely degenerate limit, and the Ti $t_{2g}$ orbitals in LaTiO$_3$ are also expected to play a role on the reduction of ordered moment.

In addition, we note that when we introduce the next-nearest neighbor transfer $t'$ in addition to the nearest neighbor transfer $t$, the AF ordered moment is strongly suppressed since $t'$ destroys the nesting of the Fermi surface. Actually, the reduction due to $t'$ is clearly shown in the recent PIRG study on the 2D Hubbard model with $t$ and $t'$.~\cite{Kashima01} When a small $t'$ of 0.2$t$ is introduced, the shape of the Fermi surface is strongly affected, and the nested structure is significantly destroyed. As a result, in the small $U/t$ region, the ground state is paramagnetic and the AFM long-range order appears only in the region of $U/t>$3.45, while for the Hubbard model with only $t$, the AFM long-range order is realized at any positive value of $U/t$ due to the perfect nesting. The ordered moment $m$ for $U/t$=4 is estimated as $\sim$0.16, which is 
considerably small relative to $m\sim$0.6 for the 2D Heisenberg model. More specifically, this value is also much smaller than $m\sim$0.35 for the 2D Hubbard model with only $t$ at $U/t$=4.~\cite{White89} This indicates that the charge fluctuations strongly reduce the AF ordered moment when the nesting of the Fermi surface is destroyed. We expect that the GdFeO$_3$-type distortion necessarily reduces the AF ordered moment through the destruction of the nested Fermi surface. Although the next-nearest neighbor transfers and the GdFeO$_3$-type distortion are relatively small in LaTiO$_3$, deviations from the nested structure together with the strong charge fluctuations can reduce the ordered moment. 

On the basis of these discussions, we conclude that the reduced ordered moment is qualitatively understood from cooperative effects of these mechanisms such as charge fluctuations, orbital degrees of freedom and the Fermi surface with imperfect nesting. The quantitative and microscopic study on the reduction is left for the future study. 

\section{Summary}

In summary, we have shown that a very simple mechanism of the La crystal field resulted from displacements of the La ions caused by the well-known GdFeO$_3$-type distortion leads to a consistent picture to dissolve the puzzling properties of the AFM(G) state in LaTiO$_3$. We have constructed the crystal field Hamiltonians by using the experimentally measured position parameters. On the basis of these Hamiltonians, we have shown that the displaced La ions in the GdFeO$_3$-type structure generates a crystal field with nearly trigonal ($D_{3d}$) symmetry. Then the threefold degenerate cubic-$t_{2g}$ levels split into three isolated levels. We note that the GdFeO$_3$-type distortion exists irrespective of the 3$d$ electronic state~\cite{Woodward97} in contrast with the Jahn-Teller distortion. The lift of the Ti 3$d$ degeneracy occurs just as a consequence of the crystal field from the La ions. In this sense, the driving mechanism proposed here is not a Jahn-Teller but a simple and new mechanism, which is inherent and universal with GdFeO$_3$-type distortion. The energies and the spin-exchange constant calculated by the effective Hamiltonian show that the lowest-orbital occupations in this crystal field stabilize the AFM(G) state, and well explains the isotropic spin-wave spectrum with considerably small spin gap. Our theory have also succeeded for the first time in explaining a number of experimental findings for LaTiO$_3$ in a unified way. The experimental results consistently accounted for by our theory include neutron scattering, NMR, resonant x-ray scattering and x-ray diffraction results.

In addition, we have also studied the effects of the $R$ crystal field in $R$TiO$_3$ with $R$ being Pr, Nd and Sm with relatively small octahedral distortions. We have also reproduced the AFM(G) state in these compounds as well as the gradual decrease of the spin-exchange constant with decreasing size of the $R$ ion, which is consistent with the decrease of $T_{\rm N}$ in experiments. Further, with increasing GdFeO$_3$-type distortion, the orbital state in the $R$ crystal field becomes to have the same symmetry as that in the ferromagnetic compounds of YTiO$_3$ and GdTiO$_3$ with large Jahn-Teller distortion, which is in agreement with resonant x-ray scattering results. In SmTiO$_3$ with moderate octahedral distortion, the $R$ crystal field and the O crystal field compete strongly, which causes a two-dimensional anisotropy in the orbital-spin structure. The depression of $T_{\rm N}$ at $R=$Sm has been attributed to this anisotropy. With decreasing size of the $R$ ion, while the $R$ crystal field dominantly determines the orbital-spin structure at $R=$La, Pr and Nd, the O crystal field becomes to dominate over the $R$ crystal field at $R=$Gd and Y through the severe competition at $R=$Sm. As observed by several experiments, although the GdFeO$_3$-type distortion is rather large, the orbital structures in GdTiO$_3$ and YTiO$_3$ with strong Jahn-Teller distortion are well characterized by the $d$-type Jahn-Teller distortion, which leads to a FM ground state.

Recently, it was theoretically predicted that spin liquid states with gapless excitation spectra are stabilized in the Mott insulators when the geometrical frustration effect is large near the Mott transition.~\cite{Kashima01,Morita02,Imada03} As compared with the spin degrees of freedom, the orbital degrees of freedom strongly couple to lattice distortions and phonons. In general, this difference makes it difficult to stabilize the orbital liquid. The present compounds provide typical examples of this tendency.

Through this study, it has been clarified that in the perovskite-type transition-metal oxides, a crystal field from the $R$ ions caused by the GdFeO$_3$-type structure lifts the $3d$ orbital degeneracy. This mechanism of lifting the orbital degeneracy competes with the usual Jahn-Teller mechanism, and eventually plays an important role in controlling orbital-spin structures. This mechanism has general importance since the GdFeO$_3$-type distortion is a universal phenomenon, which is seen in a large number of perovskite-type compounds. This mechanism may also play important roles on the electronic structures in other perovskite compounds.

\section*{Acknowledgement}
We thank N. Hamada for providing us with data on LDA calculations of LaTiO$_3$, and A. Fujimori for valuable discussions. M.M also thanks T. Mizokawa, S. Maekawa and S. Ishihara for discussions.


\begin{thebibliography}{999}
\bibitem{sec3R} Coulomb correlation effects on La $5d$ orbitals are not fully taken into account in LDA calculations, which may shift the La $5d$ level to higher energy. This shift may reduce the hybridization between La $5d$ and Ti $3d$ bands, which would reduce $H_{R2}$. Quantitative and reliable estimate of this effect is left for further studies.
\end{thebibliography}
\end{document}